\documentclass[apjfonts,appendixfloats]{emulateapj}
\usepackage{appendix,natbib}
\usepackage{amsmath}
\usepackage{courier}
\usepackage{booktabs}
\usepackage{upgreek}
\usepackage[pdfpagemode=UseNone,pdfstartview=FitH,colorlinks=true,citecolor=blue,linkcolor=blue,urlcolor=blue]{hyperref}
\usepackage[all]{hypcap}

\newcommand{\halpha}{H\ensuremath{\alpha}}
\newcommand{\hbeta}{H\ensuremath{\beta}}
\newcommand{\um}{\ensuremath{\mu}m}
\newcommand{\kmps}{km s\ensuremath{^{-1}}}
\def\msun{{\rm\,M_\odot}}
\newcommand{\loiii}{\ensuremath{L_\mathrm{[OIII]}}}
\newcommand{\lagn}{\ensuremath{L_\mathrm{AGN}}}
\newcommand{\vmax}{\ensuremath{v_\mathrm{max}}}

\begin{document}

\title{The MOSDEF survey: a census of AGN-driven ionized outflows at $\lowercase{z} = 1.4-3.8$}

\author{\sc Gene C. K. Leung\altaffilmark{1}, 
Alison L. Coil\altaffilmark{1}, 
James Aird\altaffilmark{2},
Mojegan Azadi\altaffilmark{3},
Mariska Kriek\altaffilmark{4}, 
Bahram Mobasher\altaffilmark{5}, 
Naveen Reddy\altaffilmark{5}, 
Alice Shapley\altaffilmark{6}, 
Brian Siana\altaffilmark{5},
Tara Fetherolf\altaffilmark{5}, 
Francesca M. Fornasini\altaffilmark{3},
William R. Freeman\altaffilmark{5}, 
Sedona H. Price\altaffilmark{7}, 
Ryan L. Sanders\altaffilmark{8}, 
Irene Shivaei\altaffilmark{9},
Tom Zick\altaffilmark{4}}

\altaffiltext{1}{Center for Astrophysics and Space Sciences, University of California, San Diego, La Jolla, CA 92093, USA}
\altaffiltext{2}{Department of Physics \& Astronomy, University of Leicester, University Road, Leicester LE1 7RJ, UK}
\altaffiltext{3}{Harvard-Smithsonian Center for Astrophysics, 60 Garden Street, Cambridge, MA, 02138, USA}
\altaffiltext{4}{Astronomy Department, University of California, Berkeley, CA 94720, USA}
\altaffiltext{5}{Department of Physics \& Astronomy, University of California, Riverside, CA 92521, USA}
\altaffiltext{6}{Department of Physics \& Astronomy, University of California, Los Angeles, CA 90095, USA}
\altaffiltext{7}{Max-Planck-Institut fur Extraterrestrische Physik, Postfach 1312, Garching, D-85741, Germany}
\altaffiltext{8}{Department of Physics, University of California, Davis, Davis, CA 95616, USA}
\altaffiltext{9}{Hubble Fellow, Steward Observatory, University of Arizona, 933 N Cherry Avenue, Tucson, AZ 85721, USA}

\begin{abstract}

Using data from the MOSFIRE Deep Evolution Field (MOSDEF) survey, we present a census of AGN-driven ionized outflows in a sample of 159 AGNs at $1.4 \le z \le 3.8$.
The sample spans AGN bolometric luminosities of $10^{44-47} \mathrm{~erg~s}^{-1}$ and includes both quiescent and star-forming galaxies extending across three orders of magnitude in stellar mass.
We identify and characterize outflows from the \hbeta, [OIII], \halpha ~and [NII] emission line spectra.
We detect outflows in $17\%$ of the AGNs, seven times more often than in a mass-matched sample of inactive galaxies in MOSDEF.
The outflows are fast and galaxy-wide, with velocities of $\sim 400-3500$ \kmps and spatial extents of $0.3-11.0$ kpc.
The incidence of outflows among AGNs is independent of the stellar mass of the host galaxy, with outflows detected in both star-forming and quiescent galaxies.
This suggests that outflows exist across different phases in galaxy evolution.
We investigate relations between outflow kinematic, spatial, and energetic properties and both AGN and host galaxy properties.
Our results show that AGN-driven outflows are widespread in galaxies along the star-forming main sequence.
The mass-loading factors of the outflows are typically $0.1-1$ and increase with AGN luminosity, capable of exceeding unity at $\lagn \gtrsim 10^{46} \mathrm{~erg~s}^{-1}$.
In these more luminous sources the ionized outflow alone is likely sufficient to regulate star formation, and when combined with outflowing neutral and molecular gas may be able to quench star formation in their host galaxies. 

\end{abstract}

\maketitle

\section{Introduction}
It is well established that supermassive black holes (SMBHs) are virtually ubiquitous at the centers of galaxies \citep[e.g.][]{mag98, hec14, aird18}. 
When SMBHs accrete, they are observed as active galactic nuclei (AGNs) \citep[e.g.][]{ant93, net15}.
AGNs are believed to play a crucial role in galaxy evolution as various properties of host galaxies and their SMBHs are observed to be closely connected.
Locally, the mass of the central SMBH and the velocity dispersion of the bulge of the host galaxy are found to follow a tight correlation, even though the SMBH is typically only less than one percent of the mass of the bulge \citep[e.g.][]{fer00, geb00}.
Globally, the growth rates of galaxies and SMBHs, traced by the total star formation rate (SFR) density and the total SMBH accretion density, respectively, followed a very similar trend with cosmic time, both reach a peak at the ``cosmic high noon'' of $z \sim 1-3$ \citep[e.g.][]{mad14, aird15, aza15, yang18, aird19}.

While such empirical evidence strongly suggests that the evolution of SMBHs and their host galaxies are intimately related, theoretical models of galaxy formation point to a more direct connection between AGNs and their host galaxies.
It has been observed that galaxies are inefficient in converting baryons into stars in the low ($<10^{12} M_\odot$) and high ($>10^{12} M_\odot$) mass ends of the halo mass distribution, as they have significantly lower stellar mass to halo mass ratios \citep[e.g.][]{mad96, bal12, beh13a}.
The star formation inefficiency in low mass halos is satisfactorily explained by stellar feedback \citep[e.g.][]{bou10, dut10, hop12}. 
Most galaxy formation models have to invoke AGNs to inject energy and/or momentum into the surrounding gas in high mass galaxies in order for their results to match the observed galaxy population \citep[e.g.][]{ben03, dm05, cro06, hop06a, kav17, phi18}.
Additionally, cosmological simulations have to invoke AGN feedback to produce quiescent bulge-dominated massive red galaxies that are no longer actively forming stars \citep[e.g.][]{hop08, sch15}.

AGN feedback provides an energetically feasible mechanism to remove cool gas in a galaxy or heat up cool gas to prevent further accretion of cool gas onto the galaxy to keep new stars from forming.
A commonly invoked AGN feedback process is large-scale outflows, in which an AGN produces a high velocity wind that heats or sweeps up gas over distances comparable to the size of the galaxy \citep[e.g.][]{kin11, deb12, fau12}.
This has sparked a wealth of observational studies aiming to characterize AGN outflows and test this picture.

In the local Universe ($z \lesssim 1$), outflows in AGNs have been observed to be ubiquitous and extend to kpc scales in ionized, atomic, and molecular gas \citep[e.g.][]{liu13a, liu13b, vei13, mul13, cic14, har14, zak14, mce15, woo16, rup17, per17a, per17b, min19}, though some studies suggest that outflows are more compact \citep{huse16, kar16a, kar16b, bar19a, bar19b}.
While AGN-driven outflows are found to be a widespread phenomenon at low redshift, a crucial epoch to study AGN-driven outflows is at higher redshifts of $z \sim 1-3$.
This epoch is known as the ``cosmic high noon'', when both the global SFR and AGN accretion rate are at their  peaks across cosmic time, before declining by about an order of magnitude to the present day values \citep[e.g.][]{mad14, aird15}.
Therefore, AGN-driven outflows should be most prevalent and powerful at this epoch, which is also when galaxies are undergoing peak growth.  This is therefore a crucial era to test models of AGN feedback.

Current observations at $z \sim 1-3$ of a statistical sample of AGNs with complete spectroscopic coverage of the brightest rest-frame optical emission lines are limited.
Early results at this redshift \citep[e.g.][]{nes08, can12, har12, per15, bru15} are often focused on special, extreme subclasses of AGNs such as ULIRGs or luminous quasars and were limited to small sample sizes.

More recent studies of outflows using statistical samples of AGNs at $z \sim 1-3$ have been made possible by the commissioning of multi-object near-infrared (NIR) spectrographs such as MOSFIRE \citep{mcl12} and KMOS \citep{sha13}.
The observed NIR waveband covers several important rest-frame optical emission lines, such as \hbeta, [OIII], \halpha, [NII] and [SII], at this redshift.
Of specific relevance to the study of AGN-driven outflows is the [OIII] emission line, which is a bright forbidden emission line that is traditionally used to trace the narrow-line region gas around AGNs and ionized outflows \citep[e.g.][]{ben02, sch03, mul13, zak14, woo16}.
Moreover, \hbeta, [OIII], \halpha ~and [NII] are used in emission line ratio diagnostics of the ionized gas using the BPT diagram \citep{bal81, vei87}.
NIR spectroscopic observations are therefore the ideal tool to study ionized AGN outflows at $z \sim 1-3$.

\citet{har16} study a sample of 89 AGNs up to $z = 1.7$ and detect ionized outflows in $\sim 35-50\%$ of the AGNs, indicating that outflows are quite prevalent in AGN at this epoch.
\citet{fio17} study scaling relations of AGN-driven outflows with AGN and host galaxy properties in a meta-analysis of published outflow studies spanning ionized, atomic, and molecular gas, which includes 29 ionized outflows in AGNs at $z \sim 1-3$.
Because of the nature of a meta-analysis, AGN and host galaxy properties are calculated by different authors using non-homogeneous recipes, which can affect the scaling relations obtained.
Selection biases can also exist since the sample is compiled from multiple studies.

A larger sample of ionized outflows in AGNs at $z \sim 1-3$ is presented in \citet{for18}, who study 152 AGNs found in a sample of 599 galaxies using integral field spectra (IFS) covering the \halpha, [NII], and [SII] emission lines.
While the IFS data provide two-dimensional measurements of the physical size of the outflows, the [OIII] and \hbeta ~emission lines are not covered. 
These emission lines are required to perform diagnostics using the BPT diagram.
The [OIII] forbidden line in particular provides an excellent tracer of narrow-line region gas and ionized outflows, as it is far from other bright emission lines and suffers from less contamination than the [NII] forbidden line.
More recently, \citet{coa19} study an even larger sample of luminous $z = 1-4$ quasars in detail, using the [OIII] and CIV line profiles to investigate how small scale outflows correlate with larger scale winds.

To better understand the role of AGN-driven outflows in galaxy evolution and explore their physical characteristics as a function of AGN and host galaxy properties, it is necessary to have a statistical sample of AGNs selected {\it from a uniform sample of galaxies} with complete spectroscopic coverage of the important rest-frame optical emission lines.
We achieve this using data from the recently-completed MOSFIRE Deep Evolution Field (MOSDEF) survey \citep{kri15}.
Targets in the MOSDEF survey are selected to a threshold in NIR flux, which roughly corresponds to a stellar mass limit. 
This results in a more uniform and representative census of the population of ``typical'' star-forming galaxies, as well as some quiescent galaxies.
The spectroscopic data of the MOSDEF survey provides simultaneous coverage of important optical emission lines, including the [OIII] emission line, which is particularly important in tracing the narrow-line region gas and ionized outflows around AGNs.
Furthermore, a homogeneous method is employed in this study to determine AGN and host galaxy properties.

In \citet{leung17}, we presented early results using data from the first 2 years of the MOSDEF survey and found that fast and galaxy-wide AGN-driven outflows are common at $z \sim 2$.
In this study, we take advantage of the full data set of the now completed MOSDEF survey, which contains $\sim 1500$ galaxies and $\sim 160$ AGNs.
We revisit the incidence and physical characteristics of AGN-driven outflows with a sample that almost triples that in \citet{leung17}.
This allows us to investigate in more detail the trends in the incidence of AGN-driven outflows and make more robust conclusions.
Furthermore, now equipped with a statistical sample of AGN-driven outflows, we study their physical properties as a function of AGN and host galaxy properties to better understand the impact of these outflows on their host galaxies and their role in galaxy evolution.

This paper is organized as follows.
Section \ref{data} describes the MOSDEF survey and the AGN sample.
Section \ref{results} presents methods used for detection and characterization of the outflows.
Section \ref{incidence} discusses the relation between the incidence of outflows and AGN and host galaxy properties.
Section \ref{parameters} presents the relation between the physical properties of the outflows and properties of the AGN and host galaxies.
We summarize our conclusions in Section \ref{conclusions}.

\section{Observations and AGN Sample}\label{data}
In this section we describe the dataset used in this study and the methods employed to identify AGN at various wavelengths.
We also outline how we estimate host galaxy properties using SED fitting.

\subsection{The MOSDEF Survey}

The MOSDEF survey was a four and a half year program using the MOSFIRE spectrograph \citep{mcl12} on the Keck I telescope to obtain near-infrared (rest-frame optical) spectra of $\sim 1500$ galaxies and AGNs at $1.4 \le z \le 3.8$
Targets were selected from the photometric catalogs of the 3D-HST survey \citep{skel14} in the five CANDELS fields: AEGIS, COSMOS, GOODS-N, GOODS-S and UDS.
Targets were selected by their H-band magnitudes, down to $H=24.0, 24.5$ and $25.0$ at $z=1.37-1.70, 2.09-2.61$ and $2.95-3.80$, respectively, with brighter sources and known AGNs given higher weights in targeting.
The \hbeta ~and [OIII] emission lines are covered in the J, H and K bands for the lower, middle and higher redshift intervals, respectively.
The \halpha , [NII] and [SII] emission lines are covered in the H and K bands for only the lower and middle redshift intervals, respectively.
With the full survey, 1824 spectra of galaxies and AGNs were acquired, yielding a reliable redshift for 1415 of them.
Full technical details of the MOSDEF survey can be found in \citet{kri15}.

\subsection{Emission Line Measurements}\label{line-meas}
In this section, we describe the procedures to measure emission line fluxes and kinematics in this study.
We perform the procedures described in \citet{aza17} and \citet{aza18} to simultaneously fit the \hbeta, [OIII], [NII] and \halpha ~emission lines using the \texttt{MPFIT} \citep{mar09} routine in \texttt{IDL}. 
For sources in the highest redshift interval of $2.95 \leq z \leq 3.80$, we fit only the \hbeta ~and [OIII] emission lines, as the [NII] and \halpha ~lines are not covered in the spectra.
The spectra are fitted with a continuum near the emission lines with zero slope and a maximum of three Gaussian components for each line:
\begin{enumerate}
\item A Gaussian function with FWHM $<$ 2000 \kmps, one each for the \hbeta, [OIII], [NII] and \halpha ~emission lines, representing the narrow-line emission from the AGN and/or host galaxy;
\item A broad Gaussian function with FWHM $>$ 2000 \kmps, one each for \hbeta ~and \halpha ~only, representing the broad line emission from the AGN;
\item An additional Gaussian function with FWHM $<$ 2000 \kmps , one each for the \hbeta, [OIII], [NII] and \halpha ~emission lines, representing a potential outflow component.
\end{enumerate}
The FWHM and velocity shift of each set of Gaussian functions are held equal for all of the lines of interest.
Tying the kinematics among different lines is a requirement of the restricted signal-noise ratio (S/N) in observations at high redshifts.
It assumes that the emission lines originate from the same gas, undergoing the same kinematics, which is not necessarily true.
Nonetheless, it is a common practice in studies of high-redshift galaxies and AGN due to limited S/N (see \citealt{har16} and references therein).
The flux ratios between [OIII]$\lambda 4960$ and [OIII]$\lambda 5008$ and between [NII]$\lambda 6550$ and [NII]$\lambda 6585$ are fixed at 1:3 for each Gaussian function.

We adopt the model with the narrow-line component (component 1) alone unless the addition of more components results in a decrease in $\chi^2$ corresponding to a $p$-value of $<0.01$ for the null hypothesis that the simpler model is true.
We further require that all the components have S/N greater than 3 in at least one line.

\subsection{AGN Sample}\label{agn-id}
AGNs were identified before targeting by X-ray imaging data from {\it Chandra} and/or IR imaging data from {\it Spitzer}/IRAC.
In addition, AGNs were identified in the MOSDEF spectra using rest-frame optical diagnostics.
The identification of AGNs in the MOSDEF survey is described below; full details are discussed in \citet{aza17} and \citet{leung17}.

We identified X-ray AGNs prior to the MOSDEF target selection using {\it Chandra} X-ray imaging data with depths of 160 ks in COSMOS, 2 Ms in GOODS-N, 4 Ms in GOODS-S and 800 ks in AEGIS, corresponding to hard band (2-10 keV) flux limits of $1.8\times 10^{-15}$, $2.8\times 10^{-16}$, $1.6\times 10^{-16}$ and $5.0\times 10^{-16}$ ergs s$^{-1}$ cm$^{-2}$, respectively.
The X-ray catalogs were constructed according to the prescription in \citet{lai09}, \citet{nan15} and \citet{aird15}.
The X-ray sources were matched to counterparts detected at IRAC, NIR and optical wavelengths using the likelihood ratio method \citep[e.g.][]{ bru07, luo10, aird15}, which were then matched with the 3D-HST catalogs for MOSDEF target selection.
These X-ray sources were given higher priority in MOSDEF target selection.
The 2-10 keV rest-frame X-ray luminosity was estimated by assuming a simple power-law spectrum with photon index $\Gamma = 1.9$ and Galactic absorption only.

To identify highly obscured and Compton-thick AGNs which are not as effectively selected through X-ray imaging, we also made use of MIR colors to identify AGNs in the MOSDEF survey, since high-energy photons from the AGN absorbed by dust are re-radiated at the MIR wavelengths.
We adopt the IRAC color criteria from \citet{don12}, with slight modifications, to select IR AGN using fluxes in the 3D-HST catalogs \citep{skel14}.
The criteria are:
\begin{gather} 
x={\rm log_{10}}\left( \frac{f_{\rm 5.8 \um}}{f_{\rm 3.6 \um}}\right), 
\quad y={\rm log_{10}}\left( \frac{f_{\rm 8.0 \um}}{f_{\rm 4.5 \um}}\right) \\
x \ge 0.08 \textrm{~ and ~} y \ge
0.15\\ 
y \ge (1.21\times{x})-0.27\\ 
y \le (1.21\times{x})+0.27 \label{eq:relax}\\ 
f_{\rm 4.5\um} > f_{\rm 3.6 \um} \label{eq:pl1}\\
f_{\rm 5.8 \um} > f_{\rm 4.5 \um} \label{eq:pl2}\\ 
f_{\rm 8.0 \um} > f_{\rm 5.8 \um}\label{eq:pl3}. 
\end{gather}

In our modification, we do not require Equation \ref{eq:relax} and include sources within $1 \sigma$ uncertainties of Equations \ref{eq:pl1}, \ref{eq:pl2} and \ref{eq:pl3}.
This is because the \citet{don12} criteria are designed to select luminous unobscured and obscured AGNs, but are incomplete to low-luminosity and host-dominated AGNs.
These slight modifications allow for the inclusion of a handful of additional sources close to the boundaries and are highly likely to be AGNs due to other properties \citep[see][for more details]{aza17}.

We also use emission line ratio diagnostics with the BPT diagram \citep{bal81, vei87} to identify optical AGN.
Emission line fluxes are obtained in the procedures described in Section \ref{line-meas}.
As discussed in \citet{coil15} and \citet{aza17}, the theoretical demarcation line in \citet{mel14} can satisfactorily balance contamination by star-forming galaxies and completeness of the MOSDEF AGN sample at $z \sim 2$.
The \citet{mel14} line asymptotically approaches $\log ([\mathrm{N II}] / \mathrm{\halpha}) = -0.34$ for low $\log ([\mathrm{O III}] / \mathrm{\hbeta})$.
We adopt a more restrictive criterion of $\log ([\mathrm{N II}] / \mathrm{\halpha}) > -0.3$ to identify optical AGNs in this study to reduce contamination from star formation.
In addition, we identify sources above the redshift-dependent maximum starburst demarcation line in \citet{kew13} at $z = 2.3$ as optical AGNs, as star formation alone cannot produce such high line ratios.

Among sources with a robust redshift measurement in MOSDEF, the procedures above result in 52 AGNs identified at X-ray wavelengths only and 41 AGNs identified at IR wavelengths only.
In addition, 25 sources satisfy both the X-ray and IR AGN identification criteria, resulting in a total of 118 AGNs identified in X-ray and/or IR.
79 sources are found to satisfy the optical AGN identification criteria, among which 41 were not previously identified in either X-ray or IR emission. 
This results in a total number of 159 AGNs identified at  X-ray, IR, and/or optical wavelengths.

\subsection{Stellar Masses and Star Formation Rates}\label{sed}
Stellar masses and SFR are measured by spectral energy distribution (SED) fitting using photometric data from the 3D-HST catalogs \citep{skel14} and the MOSDEF spectroscopic redshifts.
In this study, we perform SED fitting using the FAST stellar population fitting code \citep{kri09} with an additional AGN component \citep[see][]{aird18}.
We assume the FSPS stellar population synthesis model \citep{con09}, the \citet{cha03} initial mass function, a delayed exponentially declining star formation history and the \citet{cal00} dust attenuation curve.
Combined AGN templates taken from the SWIRE template library \citep{pol07} and \citet{sil04} are used \citep[see][for details]{aird18}.

\section{Outflow Detection and Characterization}\label{results}
In this section we report the methods used to detect outflows and characterize their properties, including kinematics, optical emission line ratios, physical extent, and mass and energy outflow rates in the MOSDEF dataset.

\subsection{Detection of Outflows in AGNs}\label{oflw-det}
We identify potential outflows in the AGN sample using the emission line fitting procedures described in Section \ref{line-meas}.
An outflow component is included in the best fit if the $p$-value is less than 0.01 for the null hypothesis of the absence of an outflow.
In addition, we require that both the narrow and outflow components to have S/N ratios greater than 3 to reduce false detections due to noise.
Among the 159 AGNs with a robust measurement of their redshifts, 43 host potential outflows satisfying these requirements.

We then exclude any potential outflows that have signs of on-going merger activity from our sample, as merger activity can produce similar kinematic signatures as outflows (i.e., a second kinematic component).
While outflows can be present in and triggered by mergers, we aim to measure and characterize the outflowing gas due to AGN activity in this study.
We therefore do not include mergers in our sample.
We identify mergers by visually inspecting the {\it HST} images of each source in the F160W and F606W bands.
Sources that have features suggesting double nuclei are considered potential mergers.
We define double nuclei as having two distinct peaks in brightness separated by less than $1''$ (corresponding to $\sim 8$ kpc at $z \sim 2$).
16 of the 43 potential outflows show features satisfying this criteria, and they are removed from the outflow sample.
This results in a final sample of 27 AGNs with robustly-identified outflows, corresponding to $17\%$ of a total of 159 AGNs in the full sample.
The emission line spectra with best-fit models and properties of these 27 AGNs with outflows are provided in the Appendix.
Among these 27 AGNs, 14 are identified in both X-ray and IR emission, 10 are identified in X-ray emission only, one is identified in IR emission only, and two are identified in optical emission only.

While we identify outflows from the decomposition of an asymmetric emission line profile, outflows can potentially result in single broad symmetric line profiles with widths $\gtrsim 600$ \kmps ~\citep[e.g.][]{mul13, har16, woo16}.
In this study, we only include sources with an asymmetric line profile decomposed into two Gaussian components as outflows because many of the derived quantities used in our analysis, such as velocity, spatial extent, and outflow mass, require the detection of a separate outflow component.
The outflow mass, in particular, requires measuring the luminosity of the outflow component alone.
This cannot be achieved with sources modelled with a single broad line profile.
Among the AGNs that are modelled as a well-constrained single Gaussian component, only three have a FWHM of 600 \kmps ~or greater.
These three sources with broad symmetric line profiles are not identified as outflows in the analysis below, and are very unlikely to affect the conclusions about outflow incidence in this study.

\subsection{Detection of Outflows in Inactive Galaxies}\label{gal-det}
In this study all MOSDEF targets with a reliable redshift have been fit using the emission-line fitting procedures described in Section \ref{line-meas}.
We can therefore compare the incidence of outflows detected in the emission-line spectra of AGNs and of inactive galaxies in our sample.
Out of a total of 1179 inactive galaxies with reliable redshifts, 
a significant outflow component is detected in 70 sources satisfying the spectroscopic criteria in Section \ref{oflw-det}.
After excluding potential mergers by inspecting their {\it HST} images, there are 37 inactive galaxies with outflows detected, corresponding to $3.1\%$ of the inactive galaxy sample.

To compare this result directly with the detection rate of outflows in AGNs, we need to account for the difference in the stellar mass distributions of the AGNs and galaxies in our sample.
Since AGNs at a given Eddington ratio will be more luminous in more massive galaxies, AGNs will be detected more frequently in galaxies with higher stellar masses \citep{aird12}, due to an observational selection bias.
As a result, the samples of AGNs and inactive galaxies have fairly different stellar mass distributions \citep[see][]{aza17}.
Moreover, the higher luminosities of higher mass galaxies can yield higher S/N spectra, which can potentially contribute to more frequent detection of outflows.
Therefore, it is necessary to match the stellar mass distributions of inactive galaxies to that of AGNs when comparing their outflow detection rates.

To do this, we construct histograms of the stellar mass distributions of the galaxies and AGNs, respectively, in bins of 0.25 dex.
Weights in stellar mass bins are computed by the ratio of the number of AGNs to the number of galaxies in the corresponding entries of the histograms, and are normalized to conserve the total number of galaxies.
Applying these weights to the inactive galaxy sample leads to a total of 30 outflows, corresponding to $2.5\%$ of inactive galaxies.
We also explore the effect of SFR on outflow detection in inactive galaxies by applying the same procedures above to SFR instead of stellar mass.
The weighted outflow incidence in inactive galaxies after accounting for their SFR distribution is $3.2 \%$.
This is very similar to the unweighted value, showing that the SFR distributions of the AGN and inactive galaxy sample are sufficiently alike that weighting by SFR is not necessary.

With an incidence rate of $17\%$, ionized outflows are six to seven times more frequently detected in emission in AGNs than in a mass-matched sample of inactive galaxies.
While outflows are known to be detected in absorption in star-forming galaxies at similar redshifts \citep{stei10}, here we directly compare the incidence rate of outflows detected in emission in both AGNs and inactive galaxies using the same methodology for both samples.
This factor of seven difference in outflow incidence rates between AGNs and galaxies strongly suggests that these AGN outflows are AGN-driven.
A detailed study of outflows detected as broad emission lines in inactive star-forming galaxies in MOSDEF is presented in \citet{free17}.
There is a small difference in the numbers of outflow detections between this study and  \citet{free17}, which is likely due to different constraints and criteria in the emission line fitting procedures, while the difference in outflow detection rate is primarily  driven by different selection criteria of the samples.

\begin{figure}[!htbp]
	\centering
		\includegraphics[width=0.47\textwidth]{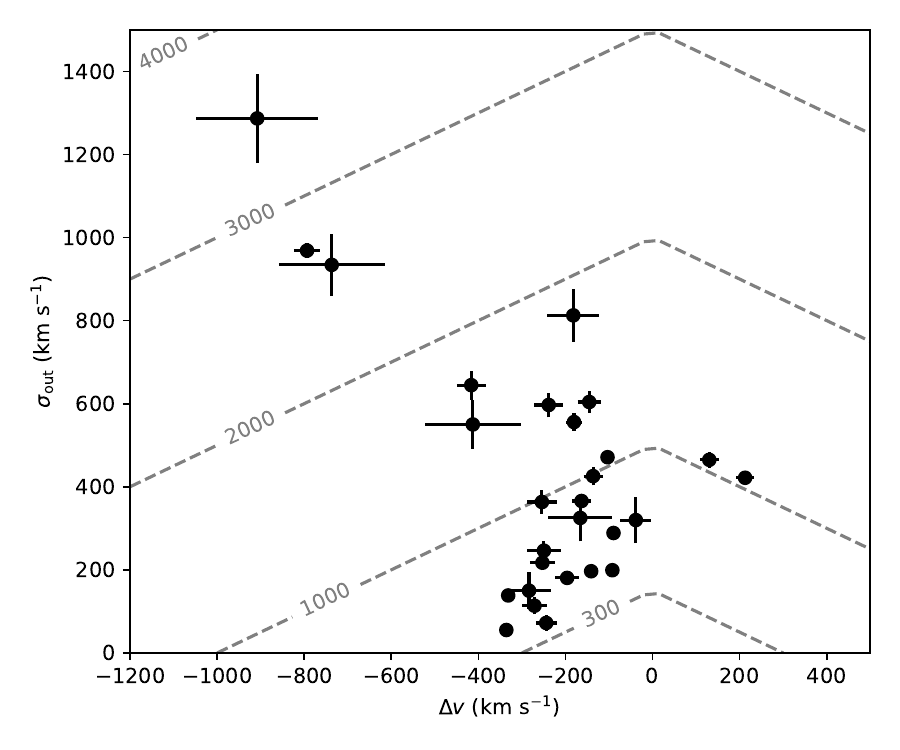}
		\caption{Velocity dispersion versus velocity shift for the detected AGN outflows in our sample. The gray dashed lines show constant $v_\mathrm{max}$ in units of km s$^{-1}$, where $v_\mathrm{max} = |\Delta v| + 2 \sigma_\mathrm{out}$ as defined in \citet{rup13}. Sources with large velocity offset have large velocity dispersions, while sources with large velocity dispersions can have small velocity offsets, resulting in the fan-shaped distribution observed in nearby AGNs \citet{woo16}.}
		\label{fig:kin}
\end{figure}

\subsection{Outflow Kinematics}\label{kin}

\begin{figure*}[!thbp]
	\centering
		\includegraphics[width=\textwidth]{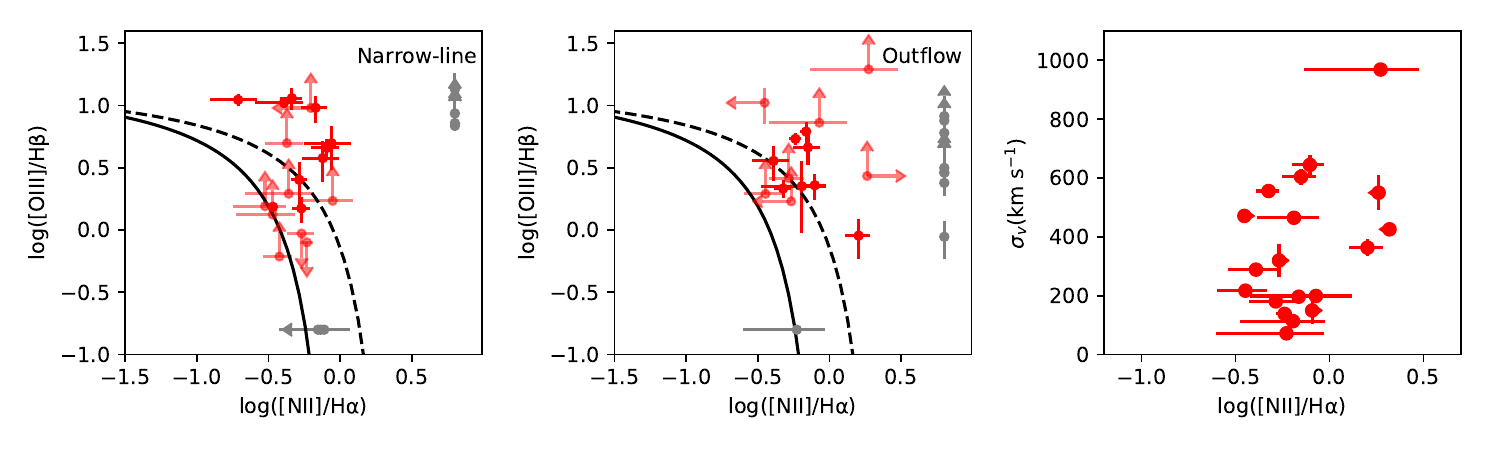}
		\caption{Left: BPT diagram for the narrow-line component of each AGN with an outflow in our sample. $3\sigma$ limits are shown when one emission line flux is not significant above $3\sigma$. Sources where only one line ratio is available are shown as gray points. Middle: Same as left panel, but for the outflow components in our sample. The line ratios of the outflow components are shifted systematically away from the star formation excitation region. The black solid and dashed lines show the \citet{kau03} and \citet{kew13} demarcation lines, respectively. Right: Velocity dispersion of the outflow component versus the [NII] to \halpha ~line ratio. If shock excitation is present, a positive correlation should be expected \citep[e.g.][]{ho14, mce15, per17b}. This is not observed in our sample, suggesting that the ouflowing gas is excited by the photoionizing radiation of the AGN rather than shocks.}
		\label{fig:BPT}
\end{figure*}

We measure kinematic parameters of the outflows in AGNs using results from the multi-component emission line fitting procedures.
One of the outflow signatures is the velocity shift ($\Delta v$) between the centroids of the outflow component and the narrow line component.
29 out of the 31 detected outflows have a negative velocity shift, meaning that they are blueshifted, while only 2 outflows are redshifted.
The magnitudes of the velocity shift range from $\sim 40 - 910$ km s$^{-1}$, with a median value of 230 km s$^{-1}$ and a median $1 \sigma$ error of 30 km s$^{-1}$.
The fact that the majority of the outflows are blueshifted is consistent with an outflow model with dust extinction \citep[e.g.][]{cren03, bae16}, where dust extinction in the galaxy reduces the flux of the receding outflowing material on the far side of the  galaxy.

We also measure the velocity dispersion of the outflow component, $\sigma_\mathrm{out}$.
The outflow velocity dispersions range from $\sim 60 - 1300$ km s$^{-1}$, with a median value of 360 km s$^{-1}$ and a median $1 \sigma$ error of 20 km s$^{-1}$.
Figure \ref{fig:kin} shows the distribution of the velocity dispersions of the outflows versus their velocity shifts.
Sources with high velocity shifts also have high velocity dispersions, while sources with relatively high velocity dispersions ($\sim$600 km s$^{-1}$) do not always have high velocity shifts.
This is similar to findings for nearby AGNs \citep{woo16}.

As both the velocity shift and velocity dispersion  account only for the component of the outflow velocity along the line of sight, outflows with a non-negligible opening angle will have a spread of observed radial velocities that are lower than the actual bulk outflow velocity.
The actual outflow velocity in three dimensions will therefore be closer to the highest or maximum radial velocity in the observed velocity distribution.
A common parameter to measure the actual velocity of the outflow is the maximum velocity ($v_\mathrm{max}$), which is defined as the velocity shift between the narrow and outflow components plus two times the dispersion of the outflow component, i.e. $v_\mathrm{max} = |\Delta v| + 2 \sigma_\mathrm{out}$ \citep{rup13}.
The maximum velocity of the outflows in our sample ranges from $\sim 400 -3500$ km s$^{-1}$, with a median value of 940 km s$^{-1}$ and a median $1 \sigma$ error of 50 km s$^{-1}$.

Another kinematic measure widely used in the literature is the non-parametric line width that contains $80\%$ of the total flux ($w_{80}$) \citep[e.g.][]{zak14}.
However, this measure depends on the relative flux ratios of the narrow and outflow components of the emission line.
While studies that make use of this measure often  have only one emission line, we simultaneously fit the \hbeta, [OIII], \halpha~and [NII] emission lines and utilize them to provide constraints on a single set of kinematic parameters.
This non-parametric measure can give rise to different results for different emission lines, due to differences in the relative flux ratios between the narrow and outflow components in different lines.
Using a single set of kinematic parameters obtained from all the available strong emission lines provides  more consistent and better constrained results for this study.
For comparison, we calculated the values of $w_{80}$ for [OIII] in the outflows in our sample, and they are highly correlated with \vmax, with $w_{80} = (0.77 \pm 0.06) \times \vmax$.

\subsection{Emission Line Ratios}\label{bpt}

Emission line ratio diagnostics, such as the BPT diagram \citep{bal81, vei87}, provide crucial information about the excitation mechanisms of the gas producing the emission.
The unique data set in the MOSDEF survey at $z \sim 2$ provides coverage for all the required optical emission lines for the BPT diagram.
Our multi-component emission line fitting procedure allows us to simultaneously measure the fluxes of both the narrow and outflow components for all the required emission lines.
Therefore, we can study the excitation mechanisms of the narrow and outflowing gas with the BPT diagram separately, allowing us to address important questions about the physical properties and impact  of these outflows.
For example, we can test the picture of 
positive AGN feedback stimulating star formation, 
if we observe increased excitation by star formation in the outflow components compared with the narrow-line components \citep{leung17}.

Figure \ref{fig:BPT} shows  [NII] BPT diagrams for the narrow-line (left) and outflow (middle) components for the AGNs with a detected outflow.
The \halpha~and \hbeta~fluxes for the narrow-line components are corrected for Balmer absorption as determined by SED modelling \citep[see][]{red15}.
Sources with S/N $< 3$ in one or both of the line ratios are shown with $3 \sigma$ limits.
Sources where only one of the two line ratios is available are shown as gray points.

The line ratios of the outflow components are shifted towards the AGN region in the BPT diagram (upper right) compared with the narrow-line components.
Three of the narrow-line component line ratios lie below the \citet{kau03} demarcation line, while all the outflow component line ratios are above this line, indicating a contribution from AGNs in the photoionization of the outflowing gas.
In addition, two more outflow component line ratios lie above the \citet{kew13} demarcation line of a  maximum starburst compared to the narrow-line component line ratios.
This trend of the outflowing gas shifting to the AGN region of this diagram is also observed in \citet{leung17}.
This shows that there is an increased contribution from AGNs to the excitation of the outflowing gas rather than from star formation.

We also consider the possibility of shock excitation by examining the relationship between the velocity dispersion and the [NII]/\halpha~ line ratio of the outflow component, as a positive correlation between the two is often interpreted as a strong tracer for shock excitation \citep[e.g.][]{ho14, mce15, per17b}.
Figure \ref{fig:BPT} (right panel) shows the distribution of velocity dispersion and the [NII]/\halpha~ line ratio for the outflows in our sample.
We compute the Spearman's rank correlation coefficient 
in this space for sources with S/N $> 3$ in both line fluxes in the outflow component, and cannot reject the null hypothesis that there is no correlation between the two quantities, with a $p$-value of 0.35 for the null hypothesis of non-correlation.
The absence of a significant correlation between velocity dispersion and the [NII]/\halpha~ line ratio suggests that as there is no evidence for shock excitation in these outflows, they are likely photoionized by the AGNs.
The increased relative contribution from AGN rather than star formation in the outflow component also suggests there is no evidence for positive AGN feedback on the galaxy scale in our sample.

\subsection{Physical Extent}\label{ext}

The physical extent of the outflows is a crucial 
measurement, needed to determine both the impact of AGN-driven outflows on the future SFR of the host galaxy and whether these outflows can expel gas over the scale of the host galaxy.
With the long-slit spectroscopic data in the MOSDEF survey, we can measure the physical extent of these AGN outflows along the slit direction.
From the 2D spectra, we create spatial profiles for the narrow-line and outflow components for the [OIII] and \halpha~ emission lines, as well as for the continuum emission.
Using results from the 1D emission line fitting, we select narrow-line or outflow dominated wavelength ranges where the narrow-line or outflow component flux is higher than the sum of all of the other components.
We also limit the wavelength ranges to be within $2 \sigma$ of the central wavelength of the respective component.
The continuum range is selected far from emission lines.
We then sum the fluxes in the 2D spectra along the wavelength axis within each wavelength range to create spatial profiles for the  narrow-line, outflow, and continuum components.
If the flux of one component is weaker than other components in all wavelengths, no spatial profile is created for the weaker component. 

To measure the physical extent of the emission in each component, we fit a Gaussian function to each of the spatial profiles.
We take the centroid of the narrow-line component as the fiducial location of the central black hole, and take the difference between the centroids of the narrow-line and outflow components, $|\Delta x|$, as the projected spatial offset between the bulk of the outflow and the central black hole.
While a broad-line component can provide an accurate measurement of the location of the central black hole \citep[e.g.][]{huse16}, most of the AGNs in our sample are Type II (with no very broad components in the \halpha ~or \hbeta ~emission lines) so this approach is not applicable here.
The centroid of the continuum component can also provide an approximate location of the center of the galaxy, but the continuum emission in our sample is substantially weaker than the narrow-line emission, providing a less accurate measurement than the narrow-line component.
If we, instead, use the centroid of the continuum to calculate the spatial offset, the resulting values only differ by 0.15 kpc on average, with a maximum of 1.05 kpc.
This difference is within the $1 \sigma$ uncertainty of $|\Delta x|$ for $79 \%$ of the sources.
Therefore, the centroid of the narrow-line component is a reliable proxy for the location of the central black hole in our sample.

\begin{figure*}[!htbp]
	\centering
		\includegraphics[width=0.9\textwidth]{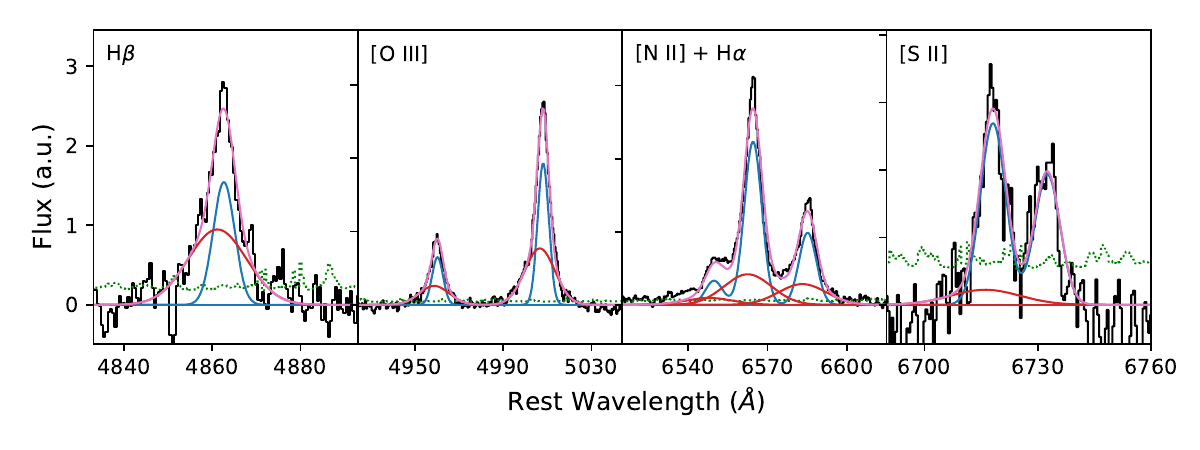}
		\caption{Stacked emission line spectrum of 23 AGNs with blueshifted outflows and no BLR emission, weighted by \vmax. The flux is shown in arbitrary units. The black line shows the observed spectrum and the green dotted line shows the error spectrum. The magenta line shows the best-fit model, while the blue and red lines show the best-fit narrow-line and outflow components, respectively. The [SII] doublet is undetected in the outflow component, with S/N of 1.1 and 0.0 for [SII]$\lambda 6716$ and [SII]$\lambda 6731$, respectively.}
		\label{fig:stack}
\end{figure*}

In the MOSDEF survey, a star was observed simultaneously with galaxies on every slitmask, providing a real-time measurement of the seeing for each target.
We deconvolve the width of the Gaussian of each spatial profile from the seeing for that slitmask in quadrature.
The deconvolved width of the spatial profile of the outflow component ($\sigma_x$) is significant (greater than three times the uncertainty) in [OIII] and/or \halpha~in 17 out of the 27 outflows in our sample.

We then combine these two measurements, the spatial offset and width of the spatial profile, to estimate the full physical extent of the outflow from the central black hole.
We define the radius of the outflow as the distance between the central black hole to the edge of the outflow where the outflow emission flux is $1/10$ the strength of the maximum, i.e. $r_{10} = |\Delta x| + 2.146\sigma_x$ or $|\Delta x| + 0.911~\mathrm{FWHM}_x$.
19 of the 27 outflows are significantly extended in  [OIII] and/or \halpha~with $r_{10}$ ranging from 0.3 to 11.0 kpc, with a median of 4.5 kpc.
We measure $r_{10}$ for both [OIII] and \halpha~as often one of the lines is impacted by a sky line.
A robust measurement of $r_{10}$ is obtained in both lines in three sources.
The spatial extents in [OIII] and \halpha ~differ by 0.1, 0.3 and 2.3 kpc in these sources, corresponding to 1.2\%, 2.8\% and 29.6\% of the radius in \halpha , respectively.

Our outflow size measurements of a few kpc are consistent with multiple studies in the local Universe \citep[e.g.][]{gre11, har14, mce15, rup17, min19} and are larger than those of some recent studies reporting outflow sizes $\lesssim 1$ kpc \citep[e.g.][]{ros18, bar19a}. 

We note that certain caveats exist with the measurement of outflow extents.
The actual outflows almost certainly have complex three dimensional structures \citep[e.g.][]{rup17} and representing the spatial extent of the outflow with a single number such as $r_{10}$ is likely incomplete.
Moreover, projection effects can on average affect the measured extent by a factor of $2/\pi \approx 0.64$.
Nonetheless, $r_{10}$ provides a lower limit to the actual spatial extent of the outflows.

\subsection{Mass and Energy Outflow Rates}\label{rates}

The mass and energy carried by the outflows are crucial parameters that determine the impact these AGN-driven outflows have on their host galaxies.
Correlations between the mass and energy outflow rates with both AGN and host galaxy properties reveal their effects on the galaxy population as a whole, as well as constrain the physical mechanisms driving the outflows.

We estimate the mass of the ionized outflowing gas by calculating the mass of recombining hydrogen atoms.
Assuming purely photoionized gas with Case B recombination with an intrinsic line ratio of  \halpha/\hbeta = 2.9 and an electron temperature of $T = 10^4$K,  following \citet{ost06} and \citet{nes17a}, the mass of the ionized gas in the outflow can be expressed as
\begin{equation}
\frac{M_\mathrm{ion}}{3.3 \times 10^8 \msun} = \left( \frac{L_{\mathrm{H}\alpha,\mathrm{out}}}{10^{43}~\mathrm{erg~s}^{-1}} \right) \left( \frac{n_e}{100~\mathrm{cm}^{-3}} \right)^{-1},
\label{eq:ha}
\end{equation}
where $L_{\mathrm{H}\alpha,\mathrm{out}}$ is the outflow \halpha ~luminosity and $n_e$ is the electron density of the ionized outflowing gas.
The electron density is taken to be $150~\mathrm{cm}^{-3}$; the rationale for this is discussed below.
Equivalently, the mass of the ionized gas in the outflow can be expressed in terms of the \hbeta ~luminosity as
\begin{equation}
\frac{M_\mathrm{ion}}{9.5 \times 10^8 \msun} = \left( \frac{L_{\mathrm{H}\beta,\mathrm{out}}}{10^{43}~\mathrm{erg~s}^{-1}} \right) \left( \frac{n_e}{100~\mathrm{cm}^{-3}} \right)^{-1},
\label{eq:hb}
\end{equation}
where $L_{\mathrm{H}\beta,\mathrm{out}}$ is the outflow \hbeta ~luminosity.

To calculate the intrinsic \halpha ~and \hbeta ~luminosity of the outflow corrected for extinction, we use the Balmer correction for the narrow line luminosity determined by SED modelling \citep[see][]{red15} and apply this correction factor to compute the intrinsic outflow luminosity, preserving the observed outflow to narrow flux ratio.
This introduces a source of uncertainty due to differential extinction in the outflowing and narrow line gas, but very likely provides a conservative lower limit to the intrinsic luminosity \citep[see][]{for18}.
For the rest of this paper we use $M_\mathrm{out} = M_\mathrm{ion}$, while noting that this is a strict lower limit as the total mass in the outflow will include molecular and neutral gas as well, which can be substantial \citep[e.g.][]{vay17, bru18, her19}.

Then the mass outflow rate is obtained by
\begin{equation}
\dot{M}_\mathrm{out} = B \frac{ M_\mathrm{out} v_\mathrm{out}}{R_\mathrm{out}}.
\end{equation} 
The value of $v_\mathrm{out}$ is taken to be \vmax ~as defined in Section \ref{kin} for the outflow velocity, while $R_\mathrm{out}$ is taken to be $r_{10}$ as defined in Section \ref{ext}.
An overall factor $B$ of $1-3$ is typically applied \citep{har18} depending on the assumed geometry of the outflows.
A spherical outflow implies a factor of 3 while an outflow covering 1/3 of the entire sphere gives a factor of 1. 
In this study, we use a fiducial value of $B=1$.
Using a different factor would simply change the overall mass and energy outflow rates by that factor.
In our sample, we find $\dot{M}_\mathrm{out} \sim 0.1 - 10^4 ~\msun~\mathrm{yr}^{-1}$, with a median of $13~\msun~\mathrm{yr}^{-1}$.

The kinetic energy outflow rate and momentum flux are then given by
\begin{equation}
\dot{E}_\mathrm{out} = \frac{1}{2} \dot{M}_\mathrm{out} v_\mathrm{out}^2
\end{equation} and
\begin{equation}
\dot{P}_\mathrm{out} = \dot{M}_\mathrm{out} v_\mathrm{out},
\end{equation}
respectively.
We find $\dot{E}_\mathrm{out} \sim 10^{40} - 10^{46} ~\mathrm{erg~s}^{-1}$, with a median of $4 \times 10^{42}~\mathrm{erg~s}^{-1}$, and $\dot{P}_\mathrm{out} \sim 10^{32}-10^{38}~\mathrm{dyn}$, with a median of $8 \times 10^{34} ~\mathrm{dyn}$.

The largest systematic uncertainty in the calculation of the mass and energy outflow rates is the electron density $n_e$.
The value of $n_e$ is difficult to measure as it typically requires detection of the outflow in a density-sensitive line such as the [SII] doublet.  
This doublet is not detected in the outflows in MOSDEF (see Section \ref{stack}).
Measurements of $n_e$ in AGN-driven outflows in the literature span a wide range of values and often have very large uncertainties.
\citet{har14} find $n_e \approx 200-1000~\mathrm{cm}^{-3}$ in a sample of $z < 0.2$ luminous AGNs, 
\citet{kak18} find spatially-resolved electron densities of $\approx 50 - 2000~\mathrm{cm}^{-3}$ in $z < 0.02$ radio AGNs, and 
\citet{rup17} find spatially-averaged electron densities of $50 - 400~\mathrm{cm}^{-3}$, with a median of $150~\mathrm{cm}^{-3}$ in a sample of $z < 0.3$ quasars.
\citet{min19} present a high resolution map of electron density in nearby AGNs with outflows and find a wide range of densities from 50 to 1000 $\mathrm{cm}^{-3}$, with a median of $250~\mathrm{cm}^{-3}$ in the outflow.
At higher redshifts, \citet{for18} report $n_e \sim 1000~\mathrm{cm}^{-3}$ in a stacked analysis of the spatially integrated spectra of AGN outflows at $z = 0.6-2.7$, while \citet{vay17} report a lower $n_e$ of $272~\mathrm{cm}^{-3}$ in the outflow of a $z = 1.5$ quasar. 
Measurements using other diagnostics yield an even wider range of values.
Studies of extended AGN scattered light infer very low densities of $< 1~\mathrm{cm}^{-3}$ \citep{zak06, gre11}.
Diagnostics with trans-auroral [SII] and [OII] emission line ratios yield densities of $10^{3-4.8}~\mathrm{cm}^{-3}$ \citep{hol11, ros18}, while SED and photoionization modelling infers density estimates of $\sim 10^{4.5}~\mathrm{cm}^{-3}$ \citep{bar19a, bar19b}.

\begin{figure*}[!htbp]
	\centering
		\includegraphics[width=0.95\textwidth]{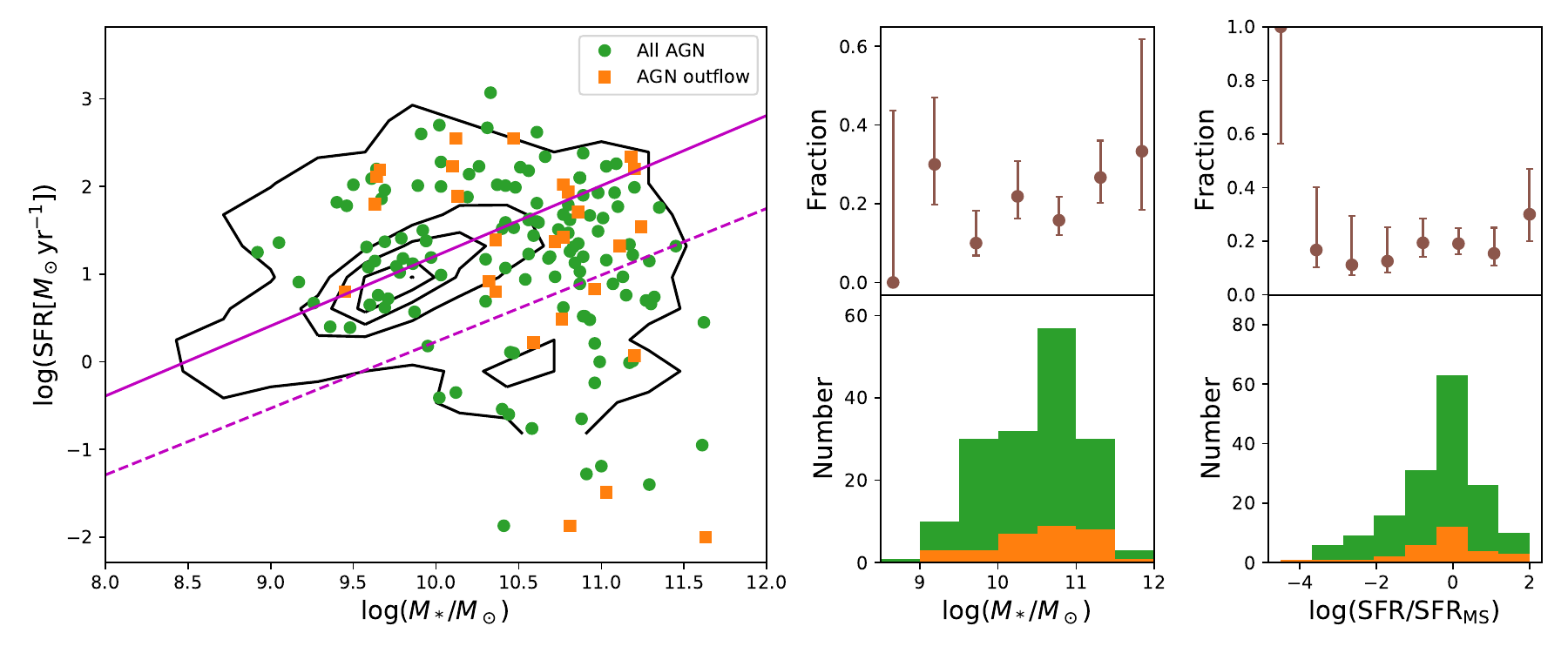}
		\caption{Left panel: Distribution of SFR and $M_*$ for all AGNs in the MOSDEF survey (blue points) and the AGNS with detected outflows in our sample (red points). Black contours show the distribution of all galaxies in the MOSDEF survey. The solid magenta line is the best-fit star-forming main sequence by \citet{shiv15}, while the dashed magenta line shows the redshift-dependent demarcation line between star-forming and quiescent galaxies by \citet{aird18}, shown here for $z=2.3$. Lower middle panel: Histogram of $M_*$ for all AGNs (green) and AGNs with outflows (orange). Lower right panel: Same as the lower middle panel, but for SFR relative to the main sequence. Upper middle panel: Fraction of AGNs that host an outflow as a function of $M_*$. Outflows are found to be distributed uniformly with $M_*$ among AGNs. However, the incidence of outflows among all galaxies can be different, since AGNs are detected more easily in higher mass galaxies (see text for details). Upper right panel: Same as the upper middle panel, but for SFR/SFR$_\mathrm{MS}$. Outflows are detected uniformly with SFR/SFR$_\mathrm{MS}$ among AGNs.}
		\label{fig:mass}
\end{figure*}

Additionally, spatially-resolved analyses of electron density show that regions of elevated electron densities ($\gtrsim 1000~\mathrm{cm}^{-3}$) are concentrated in very localized regions, while most of the other regions have much lower electron densities ($\lesssim 100~\mathrm{cm}^{-3}$, e.g. \citealt{rup17, kak18,min19}).
For instance, electron density maps of the outflowing gas in \citet{kak18} show that regions of dense gas are located in small spatial regions at radii $< 0.5$ kpc, while the electron density drops quickly with radius, reaching $< 50 ~\mathrm{cm}^{-3}$ beyond 1 kpc from the AGN.
Therefore, higher values of $n_e$ near $\sim 1000 ~\mathrm{cm}^{-3}$ reported from spatially-averaged analyses can be biased towards high density regions which dominates the total flux, as shown in \citet{min19}.
This bias can potentially lead to overestimation of the electron density over the large spatial volume of the outflows present in this study, which extend over several kpc in physical size.
While most of the high density gas in the outflows is concentrated in the most luminous compact region, substantial mass can reside in lower density regions with a much higher volume filling factors.
For instance, in a spherically symmetric mass-conserving free wind, i.e. the mass outflow rate and velocity is conserved at any radius, the density profile has to follow $n(r) \propto r^{-2}$ \citep[e.g.][]{rup05, liu13a, gen14, for18}.
In such a profile, a compact inner region dominates the density and luminosity, but the mass, which is distributed uniformly in radius, resides mostly in low density regions over a much larger volume outside this compact inner region.
Therefore, in our calculation, we take $r_{10}$ as the radius of the outflow and $n_e = 150~\mathrm{cm}^{-3}$.
$r_{10}$ represents the maximum extent of the outflow including the most extended but low surface brightness region, and the choice of $n_e$ is similar to the median electron densities in the spatially-resolved maps of \citet{rup17} and \citet{kak18}.

Furthermore, $\dot{M}_\mathrm{out}$ scales with $(n_e R_\mathrm{out})^{-1}$.
If, alternatively, a small radius and a high electron density is adopted to represent the densest compact region, one will obtain a similar outflow rate.
For example, the $0.5-1$ kpc radius of the dense compact region in \citet{kak18} is a factor of a few lower than the typical $r_{10}$ of 4.5 kpc in our outflows, while the electron density therein of $\sim 1000~\mathrm{cm}^{-3}$ is a factor of a few higher than the adopted median value of $150~\mathrm{cm}^{-3}$ here, such that the product $n_e R_\mathrm{out}$, and thus $\dot{M}_\mathrm{out}$, is largely consistent with our value.
As these two sets of values represent the outflow rates at two different radii, this result is to be expected if the mass-conserving free wind scenario is a good approximation over the range of radius concerned.
We note that outflow rate estimates based on different assumptions of the distribution of the outflowing gas can lead to different results.

\subsection{Stacked Spectrum Analysis}\label{stack}

We attempt to constrain the electron density in the outflow through measurements of the [SII] doublet using a stacking analysis of the outflow emission line spectra.
We select from our outflow sample sources with no BLR emission in \halpha ~and \hbeta ~in order to avoid contamination to the [SII] doublet.
Seven outflows are excluded by this criterion.
We also exclude two outflows with a positive $\Delta v$, i.e. a redshifted outflow component, to boost the blueshifted outflow signal in the stacked spectrum. 
These criteria result in a sample of 23 AGN outflows.
We then assign weights to each source according to the magnitude of the measured \vmax.
A stacked spectrum is constructed using the method described in \citet{shiv18}.
We perform the same line-fitting procedures described in Section \ref{line-meas} to the stacked spectrum.
The resulting stacked spectrum and its best-fit model are shown in Figure \ref{fig:stack}.
The \hbeta, [OIII], \halpha ~and [NII] emission lines all display a significant outflow component.
For the [SII] doublet, a narrow-line component is detected at S/N of 8.8 and 9.9 for [SII]$\lambda 6718$ and [SII]$\lambda 6732$, respectively.
However, the outflow component is undetected in both lines, yielding S/N of 1.1 and 0.0 in the best-fit model for the two emission [SII] lines.
No meaningful line ratios can be calculated from the line fluxes to place constraints on the electron density of these outflows.
We also construct stacked spectra using different weighting algorithms, including weighting by $\Delta v$  and outflow fluxes in [OIII] or \halpha.
The outflow component of the [SII] doublet is undetected in all of these stacked spectra.

\section{Outflow Incidence and Host Properties}\label{incidence}

In this section we study the relation between outflow incidence and the properties of the AGN or host galaxy.

\subsection{Host Galaxy Properties}\label{msfr}

\begin{figure*}[!htbp]
	\centering
		\includegraphics[width=0.9\textwidth]{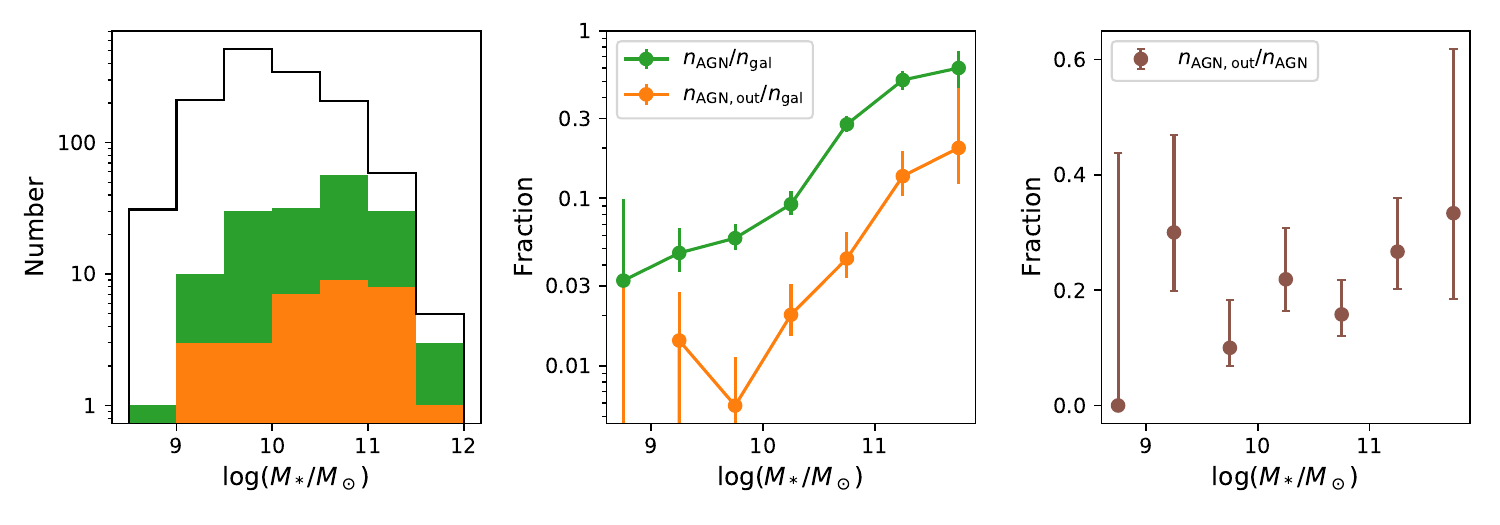}
		\caption{Left panel: Stellar mass histogram for all galaxies (black), all AGNs (green), and AGNs with outflows (orange) in the MOSDEF survey. Middle panel: The fraction of galaxies that host an AGN (green) and the fraction of galaxies that host an AGN and an outflow as a function of stellar mass. Right panel: the fraction of AGNs that host an outflow as a function of stellar mass. Both the fraction of galaxies that host an AGN and the fraction of galaxies that host an AGN and an outflow increase with $M_*$. However, the fraction of AGNs that host an outflow is uniform with $M_*$. The increasing trend seen in the middle panel is due to an observational AGN selection bias (see text for details). The right panel shows that given the presence of an AGN, the presence of an outflow is independent of stellar mass. The intrinsic incidence of AGN outflows is thus also independent of $M_*$.}
		\label{fig:mass2}
\end{figure*}

Since AGN-driven outflows are widely believed to impact the evolution of their host galaxies, especially as a form of AGN feedback appears to be needed to quench or regulate star formation in high mass galaxies, here we examine the relation between the incidence rate of AGN-driven outflows and the properties of their host galaxies.

Figure \ref{fig:mass} shows the distribution of SFR versus stellar mass for all MOSDEF galaxies (black contours), all MOSDEF AGNs (green points) and all AGNs with a detected outflow (orange points).
The star-forming main sequence for SED-based SFR in MOSDEF galaxies found by \citet{shiv15}:
\begin{equation}
\log\mathrm{SFR}[\msun ~\mathrm{yr^{-1}}] = 0.8 ~\log(M_*/M_\odot) - 6.79
\end{equation}
is shown with a solid magenta line.
We adopt the redshift-dependent minimum SFR relative to the main sequence defined in \citet{aird18}:
\begin{equation}
\begin{aligned}
\log&~\mathrm{SFR}_\mathrm{min}[\msun ~\mathrm{yr^{-1}}] \\
&= 0.76 ~\log(M_*/M_\odot) - 8.9  + 2.95~\log(1+z),
\end{aligned}
\end{equation}
where $z=2.3$, shown with a dashed magenta line.
In this study we classify galaxies that lie below this line as quiescent and galaxies above the line as star-forming.
The lower and upper right panels of Figure \ref{fig:mass} show the distributions and fractions with stellar mass and SFR, respectively, for all AGNs and AGNs with outflows in our sample.

AGNs in MOSDEF are detected in galaxies with stellar masses of $10^{8.5-12} \msun$, while outflows are detected in AGN host galaxies with $10^{9-12} \msun$.
A two-sample KS test of the stellar mass distributions of all AGNs and AGNs with outflows results in a KS statistic of 0.11 and a $p$-value of 0.89 for the null hypothesis of identical distributions.
This shows that there is no significant difference between the two distributions.
The incidence of AGN-driven outflows is independent of stellar mass, within the errors (upper middle panel).
Outflows are detected in AGNs above and below the star forming main sequence, in both star forming galaxies and quiescent galaxies.
There is no significant difference between the distributions of $\log (\mathrm{SFR}/\mathrm{SFR}_\mathrm{MS}$) (defined as the SFR relative to the main sequence) of AGNs with outflows and all AGNs, with a KS statistic of 0.11 and a $p$-value of 0.90.

It should be noted that, as discussed in \citet{leung17}, the independence of outflow incidence with SFR should not be interpreted as a {it lack} of negative AGN feedback.
The SFR in this study are obtained by SED fitting, which reflects the SFR over the past $10^8$ years \citep{ken98}.
The outflows in our sample have dynamical timescales, defined by $r_{10} / \vmax$, of $10^{5-7}$ years.
Therefore, the currently observed outflows are not expected to have an impact on the measured SFR.
The independence of outflow incidence rate with both stellar mass and SFR, shows that AGN-driven outflows are a widespread and common phenomenon that occur uniformly among galaxies both along and across the star-forming main sequence, and across different phases of galaxy evolution.

\begin{figure*}[!htbp]
	\centering
		\includegraphics[width=0.65\textwidth]{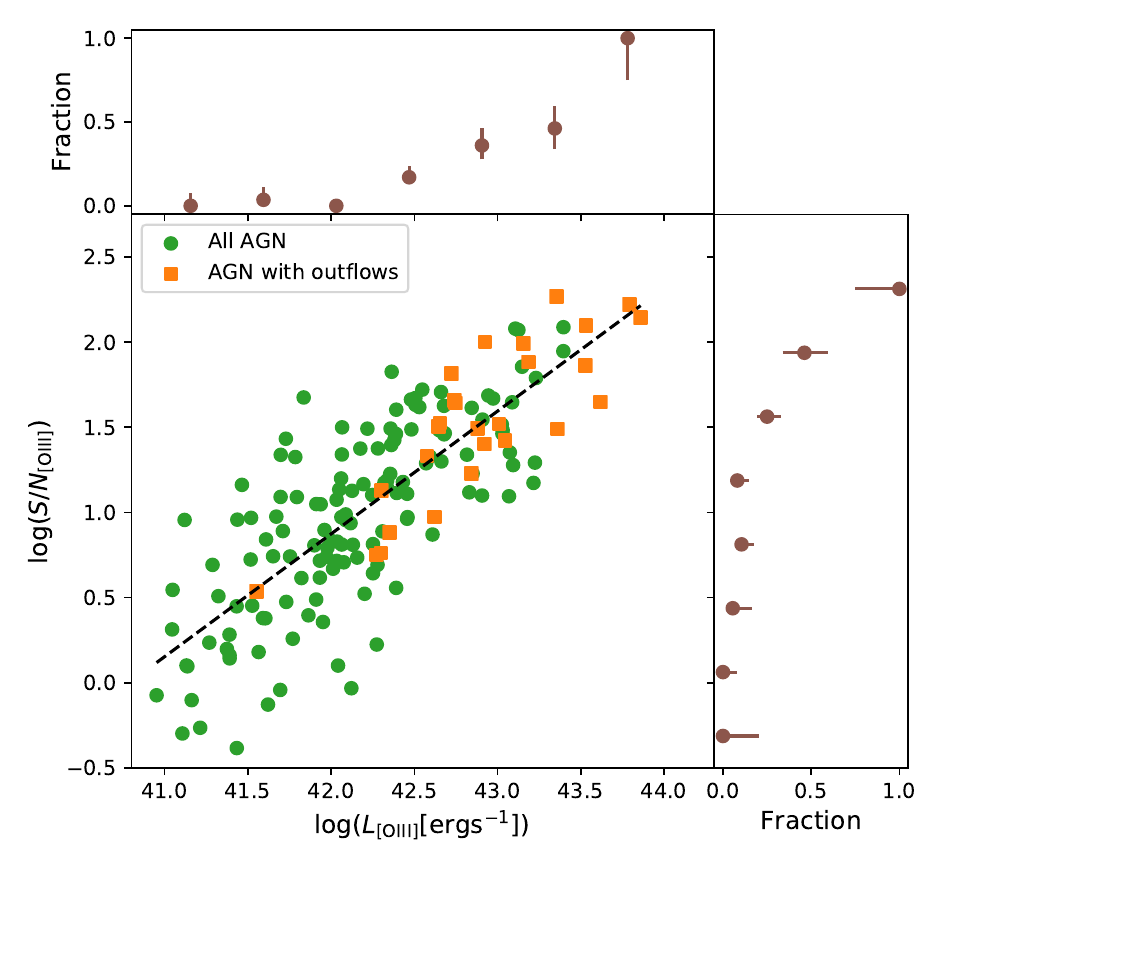}
		\caption{Distribution of signal to noise (S/N) in [OIII] and \loiii ~for all AGNs in the MOSDEF survey (green points) and AGNs with outflows (orange squares). Upper panel: The fraction of AGNs which host an outflow as a function of \loiii. Right panel: Same as the upper panel, but as a function of S/N in [OIII]. The distribution of \loiii ~and S/N$_\mathrm{[OIII]}$ are strongly correlated, and the fraction of AGNs with an outflow increases with both \loiii ~and S/N$_\mathrm{[OIII]}$. The increasing fraction cannot be independently associated with either \loiii ~or S/N$_\mathrm{[OIII]}$ in this sample.}
		\label{fig:SN}
\end{figure*}

There has been reported in the literature an increasing incidence of AGN outflows with stellar mass \citep{gen14, for18}, which appears to be contradictory to our results.
However, \citet{gen14} and \citet{for18} show that the incidence of AGN outflows {\it among all galaxies} increases with stellar mass, while here we are showing that the incidence of outflows {\it among AGNs} is independent of stellar mass.
In Figure \ref{fig:mass2}, the left panel shows the stellar mass distribution of galaxies, AGNs, and AGNs with outflows in the MOSDEF survey, while the middle panel shows the fraction of galaxies that host an AGN and the fraction of galaxies that host an AGN and an outflow.
Both of these fractions
show a strong increasing trend with stellar mass, which is consistent with the findings of \citet{gen14} and \citet{for18}.
While less than $10\%$ of galaxies host an AGN at $M_* \sim 10^9 \msun$, this fraction increases to over $60 \%$ at $M_* \sim 10^{11.5} \msun$.
This is a well-known selection effect, as AGNs of the same Eddington ratio are more luminous at higher stellar mass and thus are more likely to be detected \citep{aird12}.  
The intrinsic probability that a galaxy hosts an AGN does rise with increasing stellar mass but is not nearly as steep as the observed fraction shown here \citep{aird18}.
The fraction of galaxies that host an AGN and an outflow shows a similar increasing trend as the trend caused by AGN selection effects.
However, the fraction {\it of AGNs} that host an outflow, shown in the right panel of Figure \ref{fig:mass2}, is constant with  stellar mass, showing that given the presence of an AGN, the presence of an outflow is independent of stellar mass.
Our results indicate that after removing the selection bias of AGN identification, the underlying incidence of AGN outflows is independent of stellar mass, and that AGN-driven outflows are equally probable in galaxies across the stellar mass range probed here.

\begin{figure*}[!htbp]
	\centering
		\includegraphics[width=\textwidth]{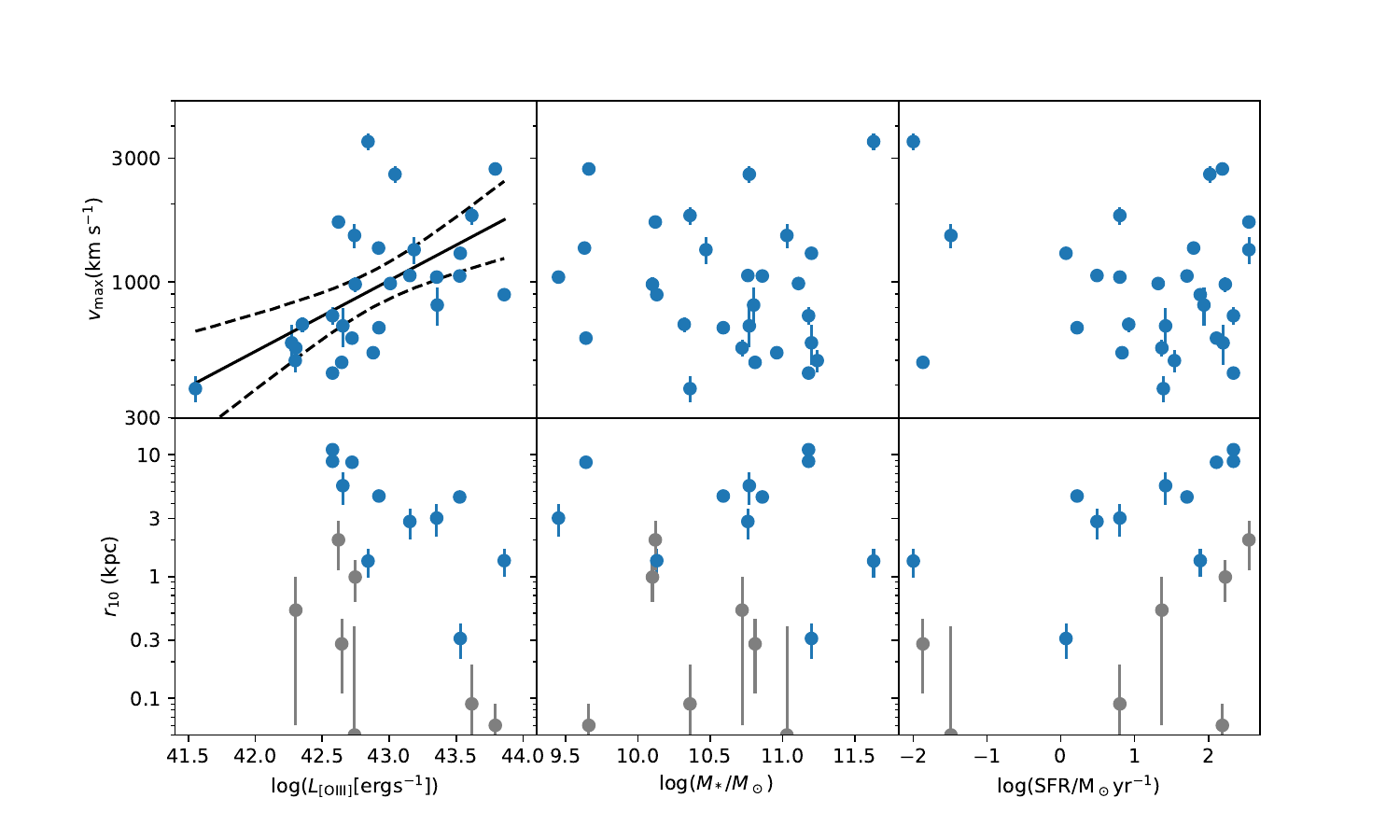}
		\caption{Top: Maximum outflow velocity (\vmax) versus [OIII] luminosity (left), galaxy stellar mass (middle) and SFR (right). Bottom: Outflow radius ($r_{10}$) versus [OIII] luminosity (left), galaxy stellar mass (middle) and SFR (right). Radius measurements with high significance (S/N $>3$) are shown in blue while those with an absolute uncertainty $<1$ kpc are shown in gray. The solid line shows the best-fit log-linear relation for significant correlations, while the dashed lines show the $90\%$ confidence intervals. \vmax ~is significantly correlated with \loiii. Although the high significance measurements of  $r_{10}$ are anti-correlated with \loiii, this is a selection effect, and this trend is absent if measurements with an absolute uncertainty $<1$ kpc are included (see text for details). There are no significant correlations with stellar mass or SFR.}
		\label{fig:vrcorr}
\end{figure*}

\subsection{[OIII] Luminosity}\label{loiii}

The [OIII] luminosity (\loiii) is a common proxy for the AGN bolometric luminosity ($\lagn$) \citep[e.g.][]{hec14}, where $\lagn$ is estimated by a constant factor times \loiii ~(a constant of 600 is found in \citealt{kau09}).
A correlation between the incidence rate of outflows and \loiii ~may therefore be interpreted as evidence that more powerful AGNs drive outflows more frequently.
However, as noted in \citet{leung17}, \loiii ~scales directly with the S/N of the [OIII] emission line (S/N$_\mathrm{[OIII]}$), which directly determines whether an outflow is detectable in emission.
We note that the AGN [OIII] luminosity in our sample is corrected for dust reddening. 
To determine the correction factor, we calculate the color excess from the Balmer decrement and combine this with the value of the MOSDEF dust attenuation curve at 5008 \AA \citep{red15}. 
This correction results in an average increase of $\sim 0.17$ dex in [OIII] luminosity for the AGNs in our sample.

Figure \ref{fig:SN} shows the the distribution of \loiii ~and S/N$_\mathrm{[OIII]}$ of all AGNs and AGNs with a detected outflow in our sample.
\loiii ~here includes emission from both the narrow-line and outflow components, which are both photoionized by the AGN.
There is an obvious correlation between \loiii ~and S/N$_\mathrm{[OIII]}$, as expected.
The fraction of AGNs with a detected outflow increases with both \loiii ~and S/N$_\mathrm{[OIII]}$.
More than $50\%$ of the AGNs have an outflow component detected at \loiii ~above $\sim 10^{43}~\mathrm{erg s}^{-1}$, which corresponds to S/N$_\mathrm{[OIII]}$ above $\sim 100$.
The fraction of AGNs with a detected outflow approaches $100\%$ at $L_\mathrm{[OIII]} \sim 10^{43.8}~\mathrm{erg s}^{-1}$ and at  S/N$_\mathrm{[OIII]} \sim 250$.
On the other hand, no AGNs have an outflow detected with S/N$_\mathrm{[OIII]}$ below 3, corresponding to $L_\mathrm{[OIII]} \sim 10^{41.5}~\mathrm{erg s}^{-1}$.
In our analysis, we require a S/N of $>3$ in both the narrow-line and outflow components individually, so any potential outflows in AGNs below this threshold are undetectable by definition.

A similar increasing trend between the fraction of AGNs with a detected outflow and \loiii ~was reported in the study of $\sim 39,000$ nearby Type II AGNs at $z < 0.3$ in \citet{woo16}.
They detect an outflow component in over $50\%$ of the AGN and composite sources at  \loiii ~above $\sim 10^{40.5}~\mathrm{erg s}^{-1}$, while the fraction approaches unity at $\sim 10^{42}~\mathrm{erg s}^{-1}$.
At the latter luminosity, AGNs in our sample at $z\sim2$ have a typical S/N$_\mathrm{[OIII]}$ of only $\sim 5$, such that detecting an outflow is challenging.
Their trend is similar to ours, except that the luminosity thresholds for outflow detection have been shifted to lower values in the lower redshift sample of \citet{woo16}.
Similar results among X-ray AGN in SDSS are also reported in \citet{per17a}.

We note that among the three sources without an outflow detection at S/N above 100 in our sample, two of them have a second kinematic component detected but are removed from our outflow sample as they are identified as potential mergers in HST imaging, following the procedures described in Section \ref{oflw-det}.
Taking this into account, a second kinematic component is detected in 7 out of 8 of the AGNs with S/N $>100$ in our sample.
As noted above, the lower overall incidence of AGN outflows we report here is in part due to the removal of potential mergers in our analysis.

In an effort to distinguish whether the trend seen in Figure \ref{fig:SN} is primarily driven by S/N or \loiii, we perform a linear regression analysis to the S/N-\loiii ~diagram.
We fit a linear function in logarithmic space to the AGNs with outflows, AGNs without outflows, and all AGNs separately, and obtained slopes of $0.75 \pm 0.11$, $0.72 \pm 0.05$ and $0.72 \pm 0.04$, respectively.
There is no difference between the distributions of AGNs with and without outflows within uncertainties.
We also use the best-fit model for all AGNs to obtain the mean \loiii ~as a function of S/N, and compare the offset of \loiii ~from the mean for AGNs with and without outflows, and find no significant difference in the distribution of \loiii ~offset between the two samples.
Our results show that \loiii ~is highly correlated with S/N in our sample, which directly affects the detection of outflows.
The effects of \loiii ~and S/N cannot be separated in the correlation between outflow incidence rate and \loiii ~seen here.

\section{Outflow Parameters and Host Galaxy Properties}\label{parameters}

With our statistical sample of AGN outflows at $z \sim 1-3$, it is possible to study the relationship between outflow properties and AGN or host galaxy properties at these high redshifts.
Such analysis can potentially reveal crucial information about the impact these outflows have on their host galaxies and help clarify their role in galaxy evolution.
As discussed in Section \ref{rates}, the calculation of the mass and energy outflow rates contains a systematic uncertainty due to the assumption of the electron density in the outflows; however, this does not affect the trends studied in this section, as all sources are affected uniformly.

\subsection{Outflow Velocity and Radius}\label{v-r}

First we study the relationship between the outflow velocity $v_\mathrm{max}$ and outflow radius $r_{10}$ and properties of the AGN and host galaxy.
In Figure \ref{fig:vrcorr}, the top row shows $v_\mathrm{max}$ against \loiii, $M_*$ and SFR.
We search for correlations between the outflow parameters and AGN or host galaxy parameters by computing the Spearman rank correlation coefficient and the associated $p$-value for the null hypothesis of non-correlation.
For correlations with a $p$-value of $<5 \%$, we perform a Bayesian linear regression analysis as described in \citet{kel07} to fit a linear function with an intrinsic scatter in logarithmic space and obtain the best-fit slope and its error.
The results are shown in Table \ref{tab:vr}.

\capstartfalse
\begin{deluxetable}{lccc}[ht!]
\centering
\tablecaption{Correlations of Outflow Velocity and Radius with AGN and Host Galaxy Properties}
\tablehead{
\colhead{Correlation} & 
\colhead{Spearman rank} &
\colhead{$p$-value} &
\colhead{Slope}
}
\startdata
\vmax vs \loiii	&	0.624	&	0.04	\%	&	0.27	$\pm$	0.09	\\
$\sigma_v$ vs \loiii	&	0.667	&	0.01	\%	&	0.41	$\pm$	0.11	\\
$\Delta v$ vs \loiii	&	0.227	&	25	\%	&		-		\\
\vmax vs $M_*$	&	-0.186	&	34	\%	&		-		\\
\vmax vs SFR	&	-0.030	&	88	\%	&		-		\\
$r_{10}$ vs \loiii	&	-0.361	&	14	\%	&		-		\\
$r_{10}$ vs $M_*$	&	0.045	&	86	\%	&		-		\\
$r_{10}$ vs SFR	&	0.447	&	6.3	\%	&		-		\\
\vmax vs $r_{10}$	&	-0.521	&	2.7	\%	&	-0.24	$\pm$	0.12	\\
$\sigma_v$ vs $r_{10}$	&	-0.569	&	1.4	\%	&	-0.33	$\pm$	0.16	\\
$\Delta v$ vs $r_{10}$	&	0.253	&	31	\%	&		-		
\enddata
\label{tab:vr}
\end{deluxetable}

The maximum velocity \vmax ~is correlated with \loiii, with a $p$-value of $0.04 \%$ and a best-fit relation of $\vmax \propto \loiii^{0.27 \pm 0.09}$.
Our results agree with the findings of \citet{fio17}, who report a common scaling between $\lagn$ and \vmax ~in an analysis of AGN-driven outflows across different phases, namely ionized, molecular, and ultrafast outflows, of $\lagn \propto \vmax ^{4.6 \pm 1.5}$, i.e. $\vmax \propto \lagn^{0.22 \pm 0.07}$.
Moreover, these results are consistent with the theoretical model for an energy conserving outflow in \citet{cos14}, which predicts a relation between $\lagn$ and \vmax ~with a power to the fifth order.
There is no significant correlation between \vmax ~and $M_*$ or SFR in our sample. 

The bottom panels of Figure \ref{fig:vrcorr} show outflow radius $r_{10}$ in [OIII] versus \loiii, $M_*$, and SFR.
The blue points show radius measurements with S/N $>3$, while the grey points show radius measurements with S/N $<3$ and an absolute uncertainty $<1$ kpc. 
The reason for showing the grey points is that while some of these measurements have fairly large relative uncertainties, if we only show measurements with S/N $>3$ (blue points), then we introduce a selection effect which results in an artificial negative correlation between outflow radius and \loiii (lower left panel).
There is a selection effect because small radius measurements with high S/N would imply a very small absolute uncertainty, which is only achievable with very high \loiii.  
It is therefore not possible to observe high S/N (blue) points in the lower left region of this plot. 
Including the grey points, which have an absolute uncertainty $<1$ kpc, we do not find a significant correlation between radius and \loiii. 
Similarly, no significant correlation is found between outflow radius and host galaxy $M_*$ or SFR.

\begin{figure*}[!htbp]
	\centering
		\includegraphics[width=\textwidth]{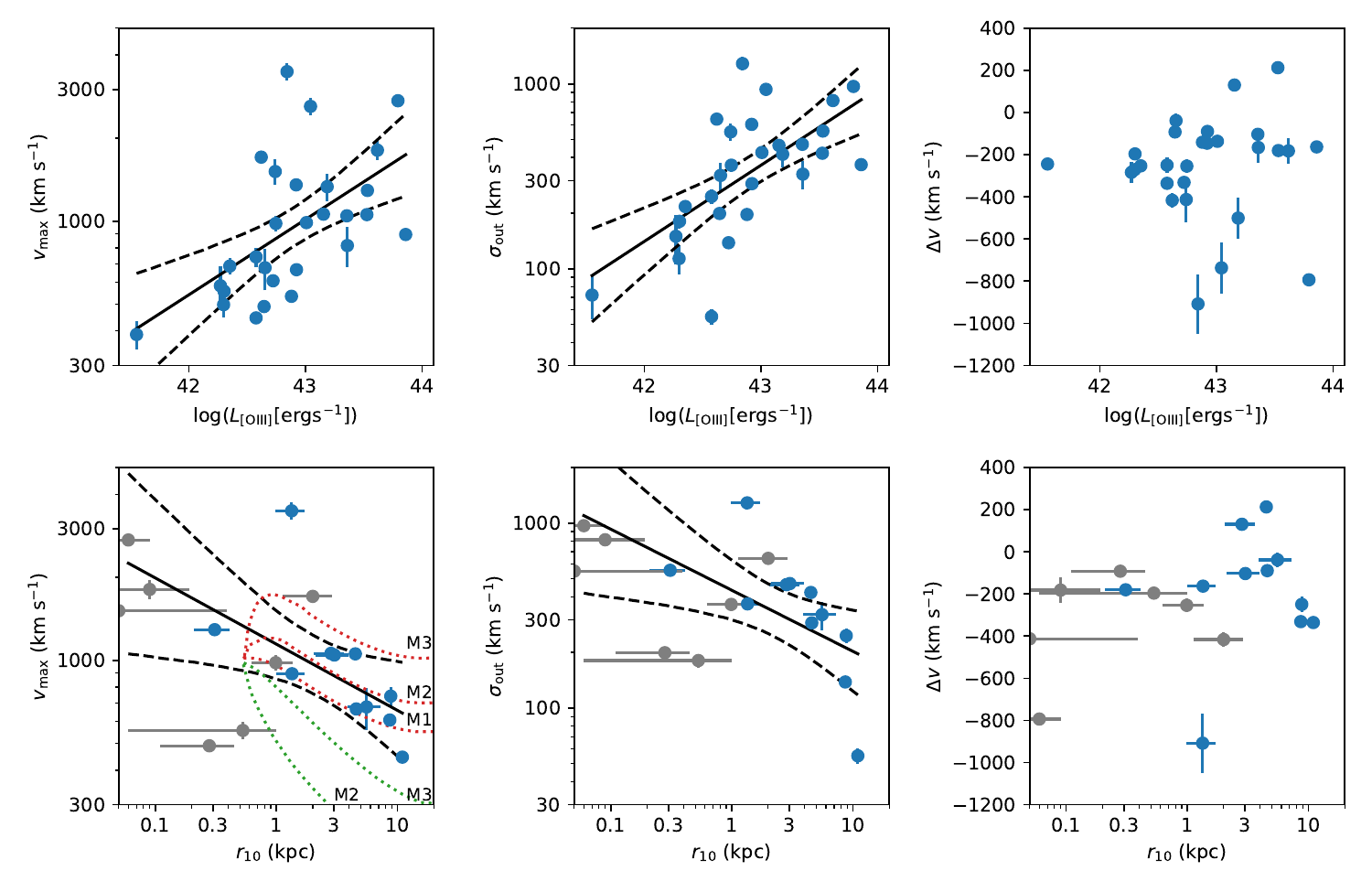}
		\caption{Top: The outflow maximum velocity (\vmax, left), velocity dispersion ($\sigma_\mathrm{out}$, middle), and velocity offset ($\Delta v$, right) versus [OIII] luminosity. Bottom: The same kinematic parameters as the top panel plotted against the outflow radius $r_{10}$. Radius measurements with S/N $>3$ are shown in blue while those with lower S/N and an absolute uncertainty $<1$ kpc are shown in grey. The solid black line is the best-fit log-linear relation, while the dashed lines show the $90\%$ confidence intervals. Both \vmax and $\sigma_v$ are correlated with \loiii, though the correlation is stronger with $\sigma_\mathrm{out}$ than with \vmax. Both \vmax and $\sigma_\mathrm{out}$ are anti-correlated with $r_{10}$. The dotted lines in the lower left panel show the analytic solutions of \citet{cos14} for energy-driven (red) and momentum-driven (green) outflows for three different black hole masses, $5\times 10^{7} \msun$ (M1), $10^8 \msun$ (M2), and $3\times 10^8 \msun$ (M3).}
		\label{fig:vlr}
\end{figure*}

Studies of AGN-driven outflows (and extended narrow-line regions) in low-redshift galaxies have revealed a positive size-luminosity relation between outflow radius and $\lagn$ \citep[e.g.][]{kang18}.
This is not observed in our sample.
However, it has also been reported that an upper limit in radius may exist above a certain luminosity, likely due to insufficient gas beyond such a radius or the over-ionization of gas, which reduces [OIII] emission \citep[e.g.][]{hain13, hain14, sun17}.
For example, \citet{sun17} report a flattening in the size-luminosity relation at $R \sim 10$ kpc and $\lagn > 10 ^{46}~ \mathrm{erg s}^{-1}$, corresponding to $\loiii \sim  10^{43}~\mathrm{erg~s}^{-1}$, while \citet{hain13} and \citet{hain14} suggest that this limit can be as low as $6-8$ kpc and $\loiii \sim 10^{42}~\mathrm{erg~s}^{-1}$.
Our sample only spans $\loiii > 10^{42}~\mathrm{erg~s}^{-1}$, 
therefore we are likely probing the flattened part of the size-luminosity relation. Our results support the existence of an upper limit in outflow size of $\sim$5-10 kpc at high luminosities.

While \vmax ~depends on \loiii, the definition of \vmax ~encompasses two different kinematic parameters, namely the velocity shift ($\Delta v$) and the outflow velocity dispersion ($\sigma_\mathrm{out}$).
We next study the relation between \loiii ~and $\Delta v$ and $\sigma_\mathrm{out}$ separately.
The results are shown in the top panels of Figure \ref{fig:vlr} and in Table \ref{tab:vr}.
We find that $\sigma_\mathrm{out}$ is significantly correlated with \loiii, with the Spearman rank correlation coefficient giving a $p$-value of $0.01 \%$. 
Interestingly, the correlation between $\sigma_\mathrm{out}$ and \loiii ~is more significant than that between \vmax ~and \loiii.
The best-fit relation is $\loiii \propto \sigma_\mathrm{out} ^ {0.41 \pm 0.11}$, which is steeper than the relation between \vmax ~and \loiii.
There is no correlation between $\Delta v$ or $|\Delta v|$ and \loiii.
We consider the absolute value of $\Delta v$ since it can have negative or positive values.
While a monotonic correlation is not observed, it does appear that only AGNs with higher \loiii ~are capable of producing higher $|\Delta v|$, while AGNs with any \loiii ~can produce small $|\Delta v|$.
Since \vmax ~is the sum of $\Delta v$ and $\sigma_\mathrm{out}$, the monotonic relation between \vmax ~and \loiii ~is mainly driven by $\sigma_\mathrm{out}$.

We also examine the relation between these outflow kinematic parameters and the outflow radius $r_{10}$ in [OIII].
The results are shown in the bottom panels of Figure \ref{fig:vlr} and in Table \ref{tab:vr}. 
With the inclusion of radius measurements with a threshold in absolute uncertainty of $<1$ kpc (grey points), there is a weak anti-correlation between radius and \vmax ~as well as radius and $\sigma_\mathrm{out}$, with the Spearman rank correlation coefficient giving a $p$-value of $2.7 \%$ and $1.4 \%$, respectively.
The best-fit relations are $\vmax \propto r_{10}^{-0.24 \pm 0.12}$ and $\sigma_\mathrm{out} \propto r_{10}^{-0.33 \pm 0.16}$.
When studying the relation between radius and \loiii, using only radius measurements with S/N $>3$ results in a significant anti-correlation which is due to a selection effect.
However, doing the same with radius and \vmax ~or $\sigma_\mathrm{out}$ ~does not lead to a significantly different result in terms of either the Spearman rank correlation coefficient or the log-linear slope.
This suggests that the weak correlation seen here between radius and \vmax ~or $\sigma_\mathrm{out}$~is unlikely due to the same selection effect, and is more likely a genuine correlation.
There is no significant correlation between radius and $\Delta v$.

\citet{bae16} study a biconical outflow model that includes dust extinction and show that $\sigma_\mathrm{out}$ is mostly driven by the intrinsic velocity and inclination of the outflow, while $\Delta v$ is primarily produced by extinction.
Using this model, our results show that either the intrinsic velocity or the inclination of the outflow, or both, is anti-correlated with the outflow radius.
The former can be explained by a deceleration of the outflowing gas as it expands, while the latter can potentially be understood as a projection effect, since the velocity is measured along the line of sight while the radius is measured perpendicular to it.
On the other hand, our data show that there is, reasonably, no correlation between extinction and outflow radius.
We explore the possibility of a deceleration of the outflowing gas as an explanation to our observed anti-correlation between radius and velocity. 
\citet{cos14} present analytic solutions to the velocity and radius of an energy- or momentum-driven outflow for three different black hole masses, $5\times 10^{7} \msun$, $10^8 \msun$, ~and $3\times 10^8 \msun$.
In Figure \ref{fig:vlr}, we overlay these analytic solutions on our observed relation between \vmax ~and radius.
The energy-driven solutions for all three black hole masses and the momentum-driven solutions for the two highest black hole masses all produce a velocity that decreases with radius between 1 and 10 kpc.
The energy-driven solutions for black hole masses of $5\times 10^{7} \msun$ ~and $10^8 \msun$ are the most consistent with the observed data, lying within the $90\%$ confidence interval of the best-fit observed relation of \vmax ~and radius.
The energy-driven solution for a black hole mass of $3\times 10^8 \msun$ ~is higher than the observed relation, while the momentum-driven solutions for all the black hole masses are lower than the observed velocity.

To summarize, in this section we explored  correlations between outflow velocity, outflow size, and AGN and  host galaxy properties.
We find that the outflow velocity is proportional to $\loiii^{0.27 \pm 0.09}$, in agreement with theoretical models of an energy-conserving outflow.
The outflow size is independent of \loiii, supporting the existence of an upper limit in outflow size of $\sim 5-10$ kpc at $\loiii > 10^{42} \mathrm{erg~s}^{-1}$.
The outflow velocity is inversely correlated with the outflow size, and the observed relation is consistent with analytic solutions of an energy-conserving outflow, while that of a momentum-conserving outflow predicts lower velocities than observed.

\subsection{Mass and Energy Outflow Rates}\label{rates-corr}

\begin{figure*}[!htbp]
	\centering
		\includegraphics[width=\textwidth]{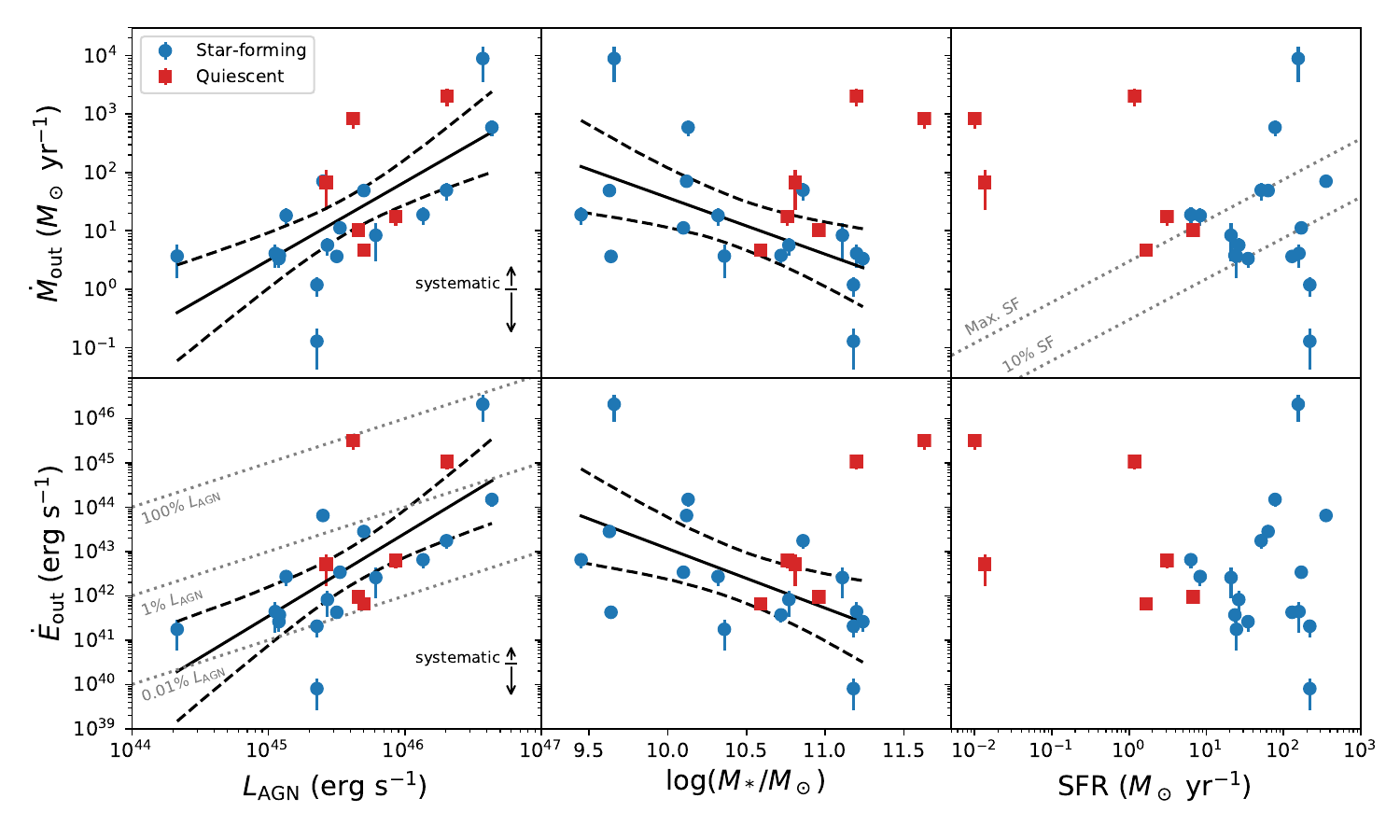}
		\caption{Top: Mass outflow rates versus $\lagn$ (left), $M_*$ (middle), and SFR (right). Star-forming galaxies are shown in blue points while quiescent galaxies are shown in red points. Bottom: Same as the top panel but for the kinetic power of the outflow. The solid lines show the best-fit log-linear slopes for significant correlations with $p$-values $<1\%$. The black arrows in the left panels show the systematic effect due to the uncertainty in the value of $n_e$, which can range from $50 - 1000 \mathrm{cm}^{-3}$. The dotted lines in the top right panel show $100\%$ and $10\%$ of the expected maximum mass-loss rate from stellar feedback by \citet{hop12}. This includes mass loss in all phases, including ionized, neutral and molecular gas, while the measurements in this study only include ionized gas. The dotted lines in the bottom left panel show the kinetic power equating $100\%$, $1\%$ and $0.01\%$ of $\lagn$. The kinetic power of all the outflows are below $100 \%$ of \lagn, and are energetically consistent with being AGN-driven. The mass outflow rates of most AGNs exceed are above $10 \%$ of expected maximum mass-loss rate from stellar feedback. Assuming $10 \%$ of the total outflow mass is ionized, these outflows cannot be produced by stellar feedback.}
		\label{fig:rates}
\end{figure*}

Figure \ref{fig:rates} shows the mass and energy outflow rates of the outflows in the sample versus $\lagn$, $M_*$, and SFR.
The error bars are the combined $1\sigma$ uncertainties from the errors on the velocity, radius, and flux.
We search for correlations between the outflow rates and $\lagn$, $M_*$, and SFR by computing the Spearman rank correlation coefficient.
\lagn ~is obtained by applying a bolometric correction factor of 600 to the total [OIII] luminosity \citep{kau09}.
For correlations with a $p$-value of $<5\%$, we fit a log-linear relation with an intrinsic scatter and obtain the best-fit slope and its error.
Six of the outflow host galaxies in our sample lie far enough below the star-forming main sequence to be classified as quiescent galaxies (see Section \ref{msfr}) and are shown as red points, while star-forming galaxies are shown as blue points.
Since quiescent galaxies have very different SFR, and therefore potentially different physical properties, from the rest of the sample, we perform the correlation analysis on the entire outflow sample containing both star-forming and quiescent galaxies, as well as on the sub-sample of outflows in star-forming galaxies only.
Table \ref{tab:rates} shows the Spearman rank correlation coefficients, $p$-value, and the best-fit slope in the case of a $p$-value $<5\%$.

Both the mass outflow rate and energy outflow rate are significantly correlated with the AGN bolometric luminosity, with $p$-values of $0.05\%$ and $0.006\%$, respectively.
The best-fit relations are $\dot{M}_\mathrm{out} \propto \lagn^{1.34 \pm 0.37}$ and $\dot{E}_\mathrm{out} \propto \lagn^{1.87 \pm 0.51}$.
The typical value of $\dot{E}_\mathrm{out}$ is between $0.01-1 \%$ of $\lagn$, with a median of $0.1\%$.
The log-linear slope of $\dot{E}_\mathrm{out}$ versus $\lagn$ is $>1$, meaning outflows in higher luminosity AGNs have a higher kinetic coupling efficiency.
This finding is consistent with that of \citet{fio17}, who find a somewhat lower but greater than unity log-normal slope of $1.29 \pm 0.38$ in ionized outflows in AGNs.
Outflows in both star-forming and quiescent galaxies follow the same trend with $\lagn$.
This shows that at a given $\lagn$ outflows are not significantly more powerful in either quiescent or star-forming galaxies.

\capstartfalse
\begin{deluxetable*}{lcccccc}[!b]
\centering
\tablecaption{Correlations of Outflow Parameters with AGN and Host Galaxy Properties}
\tablehead{
\colhead{} & 
\multicolumn{3}{c}{Star-forming and quiescent galaxies} &
\multicolumn{3}{c}{Star-forming galaxies only}
\\
\cmidrule(lr){2-4} \cmidrule(lr){5-7}
\colhead{Correlation} & 
\colhead{Spearman rank} &
\colhead{$p$-value} &
\colhead{Slope} &
\colhead{Spearman rank} &
\colhead{$p$-value} &
\colhead{Slope}
}
\startdata
$\dot{M}_\mathrm{out}$ vs $L_\mathrm{AGN}$	&	0.666	&	0.053	\%	&	1.34	$\pm$	0.37	&	0.703	&	0.17	\%	&	1.26	$\pm$	0.38	\\
$\dot{E}_\mathrm{out}$ vs $L_\mathrm{AGN}$	&	0.735	&	0.006	\%	&	1.87	$\pm$	0.51	&	0.765	&	0.034	\%	&	1.78	$\pm$	0.50	\\
$\dot{M}_\mathrm{out}$ vs $M_*$	&	-0.217	&	32	\%	&		-		&	-0.575	&	1.57	\%	&	-0.97	$\pm$	0.41	\\
$\dot{E}_\mathrm{out}$ vs $M_*$	&	-0.242	&	27	\%	&		-		&	-0.606	&	1.00	\%	&	-1.34	$\pm$	0.56	\\
$\dot{M}_\mathrm{out}$ vs SFR	&	-0.303	&	16	\%	&		-		&	-0.070	&	79	\%	&		-		\\
$\dot{E}_\mathrm{out}$ vs SFR	&	-0.236	&	28	\%	&		-		&	0.025	&	93	\%	&		-		\\
$\eta$ vs $L_\mathrm{AGN}$	&	0.655	&	0.070	\%	&	1.83	$\pm$	0.94	&	0.658	&	0.41	\%	&	1.46	$\pm$	0.55	\\
$\eta$ vs $M_*$	&	-0.067	&	76	\%	&		-		&	-0.514	&	3.5	\%	&	-1.08	$\pm$	0.51	\\
$\eta$ vs SFR	&	-0.723	&	0.010	\%	&	-1.35	$\pm$	0.21	&	-0.499	&	4.1	\%	&	-1.09	$\pm$	0.70	\\
$\widetilde{\eta}$ vs $L_\mathrm{AGN}$	&	0.641	&	0.10	\%	&	1.89	$\pm$	0.50	&	0.690	&	0.22	\%	&	1.93	$\pm$	0.60	\\
$\widetilde{\eta}$ vs $M_*$	&	-0.507	&	1.35	\%	&	-1.19	$\pm$	0.46	&	-0.798	&	0.012	\%	&	-1.84	$\pm$	0.44	\\
$\widetilde{\eta}$ vs SFR	&	-0.173	&	43	\%	&		-		&	-0.075	&	78	\%	&		-		
\enddata
\label{tab:rates}

\end{deluxetable*}

There is no significant correlation between mass or energy outflow rate and stellar mass for the full sample, which includes both star-forming and quiescent galaxies.
However, in the sub-sample of star-forming galaxies only, there is a marginally significant negative correlation between $\dot{M}_\mathrm{out}$ and $M_*$ with a $p$-value of $1.57\%$.
We note that this correlation could potentially be driven by two outlying points with the highest and lowest mass outflow rates.
If we remove those two points, we obtain a weak negative correlation with a $p$-value of $6.4 \%$.
There is also a negative correlation between $\dot{E}_\mathrm{out}$ and $M_*$ with a $p$-value of $1.0 \%$. 
The best-fit relations for star-forming galaxies are $\dot{M}_\mathrm{out} \propto M_*^{-0.97 \pm 0.41}$ and $\dot{E}_\mathrm{out} \propto M_*^{-1.34 \pm 0.56}$.
If these two correlations are robust, the negative slope indicates that along the star-forming main sequence, higher mass galaxies produce less powerful AGN outflows.
On the other hand, this trend is not observed in quiescent galaxies, where higher mass galaxies appear to have more powerful outflows at a given stellar mass.
However, the small number of outflows in quiescent galaxies in our sample prevents the detection of any potential correlations.
This correlation could imply that at higher stellar masses ($M_* > 10^{10.5} \msun$), quiescent galaxies host more powerful AGN-driven outflows than star-forming galaxies.
There is no significant correlation between the outflow rates and SFR of the host galaxy.

\subsection{Physical Driver of the Outflows}\label{driver}

Our measurements of mass and energy outflow rates allow us to constrain the physical drivers of these outflows.
For example, the ratio of $\dot{E}_\mathrm{out}/\lagn$ compares the kinetic energy carried by the outflows to the bolometric luminosity of the AGN.
For all of the outflows in our sample, $\dot{E}_\mathrm{out}$ is less than $100 \%$ of $\lagn$, while most are between $0.01 - 1 \%$ of $\lagn$, meaning that the AGN is energetically sufficient to drive these outflows.

On the other hand, another possible driver of galactic outflows is stellar feedback.
\citet{hop12} estimate a mass loss rate in outflows driven by stellar feedback as a function of SFR given by
\begin{equation}
\dot{M}_\mathrm{stellar, max} = 3 \msun ~\mathrm{yr}^{-1} \left( \frac{\mathrm{SFR}}{\msun ~\mathrm{yr}^{-1}} \right)^{0.7}.
\end{equation}
This mass loss rate includes not only direct mass loss from supernovae and stellar winds but also the subsequent entrainment of the interstellar medium.
Moreover, it includes all material that is being ejected out of the galaxy, in all phases, locations, and directions, which will generally be larger than what is directly observed.
Therefore it can be considered as an upper limit on the observed mass outflow rates that could be due to stellar feedback alone.
This maximum mass loss rate as a function of SFR is shown in the upper right panel of Figure \ref{fig:rates} in the gray dashed line.

Nine of the outflows have mass outflow rates exceeding the predicted maximum mass loss rate from stellar feedback, meaning that stellar feedback is decidedly insufficient to drive these ten outflows.
Moreover, it should be noted that the observed $\dot{M}_\mathrm{out}$ here only includes ionized outflows, while $\dot{M}_\mathrm{stellar, max}$ accounts for mass loss in all phases, such that more sources could potentially have a total mass loss rate greater than $\dot{M}_\mathrm{stellar, max}$. 
In particular, the molecular gas mass can easily be an order of magnitude greater than the ionized gas mass in AGN outflows \citep[e.g.][]{her19}.
17 of the 22 outflows have mass outflow rates exceeding $10\%$ of the expected maximum mass-loss rate from stellar feedback.
Assuming ionized gas makes up $10 \%$ of the total outflowing mass, the majority of the outflows in our sample cannot be driven by stellar feedback.
Combined with the fact that the kinetic power of all the outflows is less than \lagn, the outflows lie in the AGN region in the BPT diagram, and the seven-fold increase in outflow incidence among AGNs relative to inactive galaxies, AGN are the likely drivers of the outflows in our sample.

\begin{figure*}[!htbp]
	\centering
		\includegraphics[width=\textwidth]{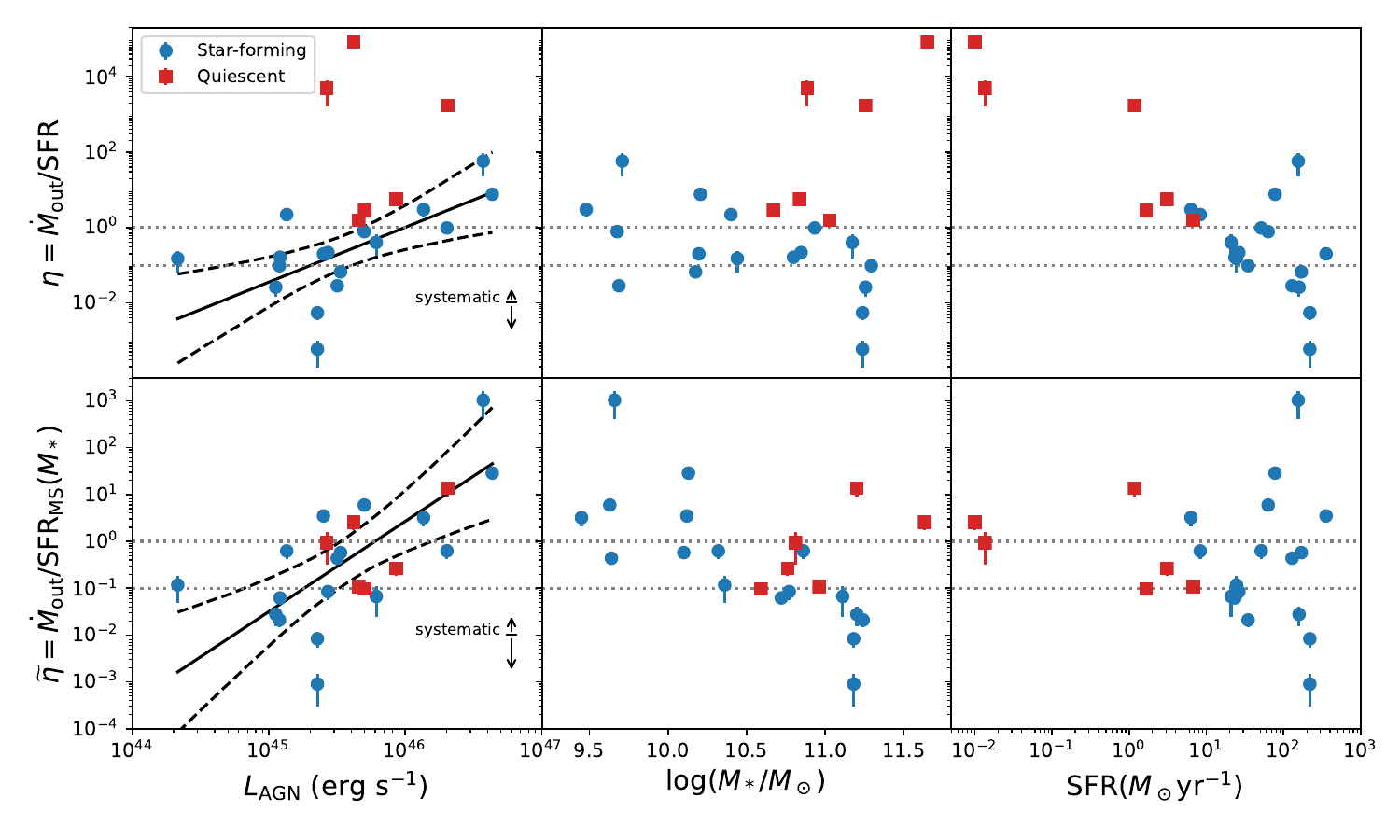}
		\caption{Top: Mass-loading factors of the outflows versus $\lagn$ (left), $M_*$ (middle), and SFR (right). Star-forming galaxies are shown in blue points while quiescent galaxies are shown in red points. Bottom: Same as the top panel but for a re-scaled mass-loading factor $\widetilde{\eta} = \dot{M}_\mathrm{out} / \mathrm{SFR}_\mathrm{MS}$, which represents the mass-loading factor of the outflow if the galaxy is on the star-forming main sequence. The dotted lines show mass-loading factors of unity and 0.1. The mass-loading factors are greater than 0.1 for most of the outflows. If $10\%$ of the outflowing mass is ionized, these outflows would have significant impact on the star formation of the host galaxy. The re-scaled mass-loading factor of quiescent galaxies are $\gtrsim 0.1$, showing that the current outflows would be sufficient to regulate the past SFR of the host galaxy.}
		\label{fig:eta}
\end{figure*}

\subsection{Mass Loading Factor}\label{eta}

To quantify the potential impact of the outflows in the context of their host galaxies, it is common to calculate the mass-loading factor of the outflows, defined as $\eta = \dot{M}_\mathrm{out}/\mathrm{SFR}$.
The mass-loading factors of outflows in our sample span a wide range from $\sim 10^{-3} - 10^5$, with a median of $0.8$.
Outflows residing in quiescent galaxies have systematically elevated mass-loading factors due to their very small SFRs and the fact that $\eta$ is defined as $\dot{M}_\mathrm{out} / \mathrm{SFR}$.
Among the outflows in star-forming galaxies, the maximum mass-loading factor is $\sim 50$, and the median is $\approx 0.2$.

Figure \ref{fig:eta} (upper panel) shows the mass-loading factors of the outflows versus $\lagn$, $M_*$, and SFR.
A strong correlation is observed in $\eta$ versus $\lagn$ with a $p$-value of $0.07 \%$ for all galaxies and $0.41\%$ for star-forming galaxies only.
While the mass-loading factor in star-forming galaxies follows a tight correlation with $\eta \propto \lagn^{1.46 \pm 0.55}$, the  values in quiescent galaxies deviate systematically upward of this trend, and the best-fit relation for quiescent and star-forming galaxies combined is $\eta \propto \lagn^{1.83 \pm 0.94}$.
This is due to the systematically lower SFR of quiescent galaxies which elevate their mass-loading factors.
In star-forming galaxies, the mass-loading factor reveals the impact of the outflows on the host galaxy while it is still in the process of forming stars, and can therefore provide information on whether the outflow can quench or regulate star formation. 
However, in quiescent galaxies the mass-loading factor reveals the impact of the outflows after star formation in the host galaxies has been quenched, and therefore can only indicate whether the outflow can keep the host galaxy quenched.
Therefore, the correlation observed in star-forming galaxies alone is more relevant to the potential role of the outflows in the initial quenching of star formation in the host galaxy.

There is a weak negative correlation in star-forming galaxies between $\eta$ and $M_*$, with a $p$-value of $3.5 \%$.
The best-fit relation is $\eta \propto M_*^{-1.08 \pm 0.51}$.
There is no correlation between $\eta$ and $M_*$ when combining star-forming and quiescent galaxies.
This is similar to the trend observed between $\dot{M}_\mathrm{out}$ and $M_*$.
However, the correlation in $\eta$ is less significant, and the slope is somewhat lower, though it is within the $1\sigma$ uncertainty of the slope with $\dot{M}_\mathrm{out}$.
\citet{fio17} also find a weak negative correlation between the mass-loading factor and stellar mass in outflows in their sample.

\begin{figure*}[!htbp]
	\centering
		\includegraphics[width=0.7\textwidth]{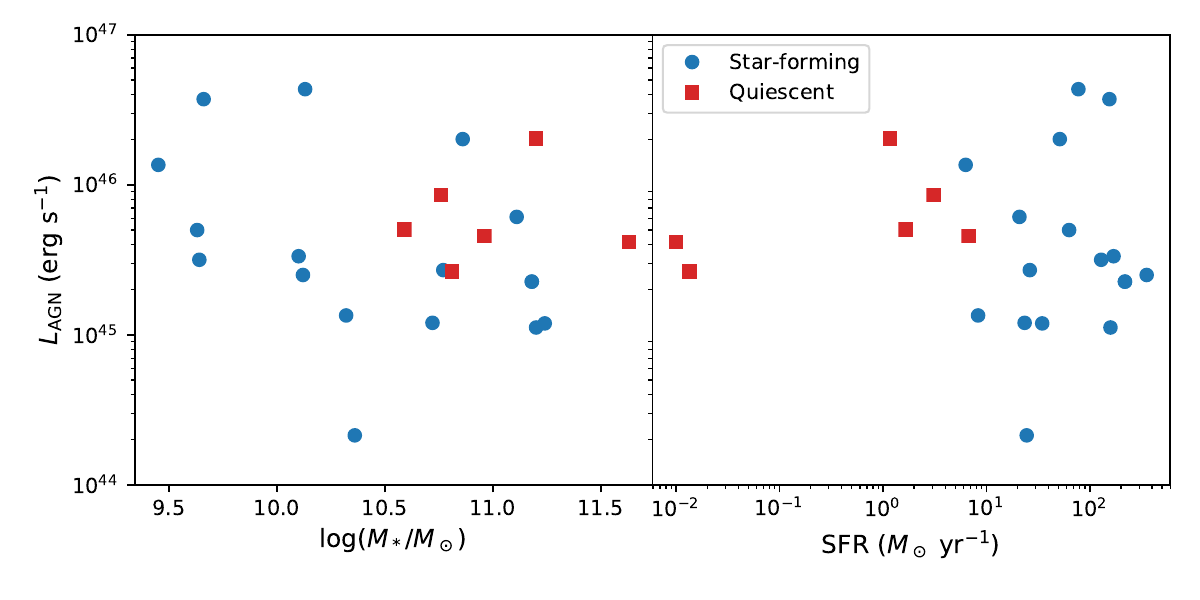}
		\caption{Left: AGN bolometric luminosity versus stellar mass. Right: AGN bolometric lumiosity versus SFR. Outflows in star-forming galaxies are shown in blue points, while outflows in quiescent galaxies are shown in red points. There is no significant correlation between $\lagn$ and $M_*$ or SFR in our outflow sample.}
		\label{fig:LAGN}
\end{figure*}

A strong artifact is observed in the plot of $\eta$ and SFR, as $\eta$ is proportional to $1/\mathrm{SFR}$ by definition.
This artifact makes it apparent that $\eta$ is heavily affected by the SFR of the host galaxy, and any interpretation of correlations between $\eta$ and other quantities must take into account any potential underlying correlation with SFR.  In particular, this makes it clear that star-forming and quiescent galaxies should generally be separated in such analyses, which is not often done.
Figure \ref{fig:LAGN} shows the distribution of $\lagn$ versus $M_*$ and SFR for our outflow sample.
We do not observe any correlations between $\lagn$ and either $M_*$ or SFR.
There is also no correlation between $M_*$ and SFR among the outflow host galaxies in our sample (see Figure \ref{fig:mass}).
Therefore, the correlations between $\eta$ and $\lagn$ and $M_*$ are not driven by underlying correlations with SFR.

For quiescent galaxies, the mass-loading factor is relevant for understanding whether the outflow might keep star formation quenched, but it does not indicate whether the outflows could have initially quenched star formation, since it compares the mass outflow rate with the SFR after quenching has already occurred.
To answer the question of whether these outflows could have quenched the star formation initially in these galaxies, it is informative to compare the outflow rate with an approximate past SFR of the host galaxy while it was still forming stars at a high rate.
To do this we use the stellar mass of the galaxy and the corresponding SFR of the star-forming main sequence.
We define a re-scaled mass-loading factor $\widetilde{\eta}$ as $\dot{M}_\mathrm{out} / \mathrm{SFR}_\mathrm{MS}$, where $\mathrm{SFR}_\mathrm{MS}$ is the SFR the galaxy would have if it was on the star-forming main sequence, given the current stellar mass of the galaxy.
This effectively compares the outflow rate with the stellar mass of the host galaxy instead of the current SFR.
Such a definition can also eliminate the artifact between $\eta$ and SFR mentioned above.
The lower panel of Figure \ref{fig:eta} shows the re-scaled mass-loading factor versus $\lagn$, $M_*$, and SFR.

The re-scaled mass-loading factors are no longer elevated for quiescent galaxies.
The re-scaled mass-loading factor is above 0.1 for the majority of the outflows, and has a median of 0.4. 
The values of $\widetilde{\eta}$ of quiescent galaxies fall along the main trend observed for star-forming galaxies.
A correlation between the re-scaled mass-loading factor and AGN luminosity is observed at a $p$-value of $0.10 \%$ for all galaxies and $0.22 \%$ for star-forming galaxies only.
An analysis combining star-forming and quiescent galaxies yields a best-fit relation of $\widetilde{\eta} \propto \lagn^{1.89 \pm 0.50}$, consistent with that of star-forming galaxies only, which follows a power of $1.93 \pm 0.60$.

The artificial correlation between $\eta$ and SFR is eliminated with the re-scaled definition of $\widetilde{\eta}$.
However, this introduces another artificial negative correlation between $\widetilde{\eta}$ and $M_*$, as the re-scaled mass-loading factor is proportional to the inverse of $\mathrm{SFR}_\mathrm{MS}$, which is proportional to $M_*$ by definition.
This highlights that extra caution is necessary when interpreting correlations between mass-loading factors and host galaxy properties.

On the other hand, the correlation between mass-loading factor and $\lagn$ is significant both in the standard definition ($\eta$) in star-forming galaxies and in the re-scaled definition ($\widetilde{\eta}$) in star-forming and quiescent galaxies.
This indicates that the correlation between mass-loading factor and $\lagn$ is a truly physical correlation and is not due to systematic effects. 
More luminous AGN therefore drive outflows that have greater potential to impact their host galaxy.

The mass-loading factors in ionized gas are greater than 0.1 for 11 out of 17 of the outflows in star forming galaxies. 
Studies of AGN-driven molecular outflows show that the mass outflow rate in the molecular phase can be comparable to or an order of magnitude greater than that in the ionized phase in $z \sim 2$ \citep{vay17, bru18, her19} AGNs.
If $10\%$ of the outflowing mass is ionized, the combined ionized and molecular outflows in these systems would have significant impact on the star formation of the host galaxy.
The mass loading factor in star-forming galaxies is positively correlated with \lagn, and is capable of exceeding unity at $\lagn \gtrsim 10^{46} \mathrm{~erg~s}^{-1}$. 
This shows that more luminous AGNs drive more powerful outflows that have higher impact to the host star-forming galaxy.
For outflows in quiescent galaxies, the re-scaled mass loading factor is greater than 0.1 in four out of six detected ionized outflows.
This shows that the current outflows in these quiescent galaxies would likely have been  sufficient to regulate the past SFR of the host galaxy.
Moreover, the current outflows in these quiescent galaxies have mass-loading factor greater than unity, implying that they are capable of keeping star formation quenched in these galaxies.
We note that some caution should be exercised when interpreting the impacts of these outflows, considering systematic uncertainties in the outflow rates as discussed above.

\subsection{Momentum flux}\label{mom}

The momentum flux carried by the outflows can provide useful information for models of AGN-driven winds.
Two common models of large-scale AGN-driven winds are momentum-conserving winds driven by radiation pressure on dust  \citep[e.g.][]{thom15, cos18} and energy-conserving winds driven by fast, small-scale winds \citep[e.g.][]{fau12, cos14}.
According to the radiation pressure-driven models, the momentum flux of the outflows is given by
\begin{equation}
    \dot{P}_\mathrm{out} = \uptau_\mathrm{IR} \frac{\lagn}{c},
\end{equation}
where $\uptau_\mathrm{IR}$ is the optical depth in IR.
For energy-driven models, the fast, small-scale winds transport a momentum comparable to $\lagn/c$, and do work to the surrounding material, increasing the momentum of the large-scale wind.
Assuming half of the kinetic energy of the small-scale wind is transferred to the large-scale wind \citep{fau12}, 
\begin{equation}
    \frac{\dot{P}_\mathrm{out}}{\dot{P}_\mathrm{in}} \approx \frac{1}{2} \frac{v_\mathrm{in}}{v_\mathrm{out}},
\end{equation}
and thus
\begin{equation}
    \dot{P}_\mathrm{out} \approx \frac{1}{2} \frac{v_\mathrm{in}}{v_\mathrm{out}} \frac{\lagn}{c},
\end{equation}
where $\dot{P}_\mathrm{in} = \lagn/c$ and $v_\mathrm{in}$ are the momentum flux and velocity of the small-scale wind, respectively, and $v_\mathrm{out}$ is the velocity of large-scale wind constituted by the swept up material.

\begin{figure}[!htbp]
	\centering	\includegraphics[width=0.47\textwidth]{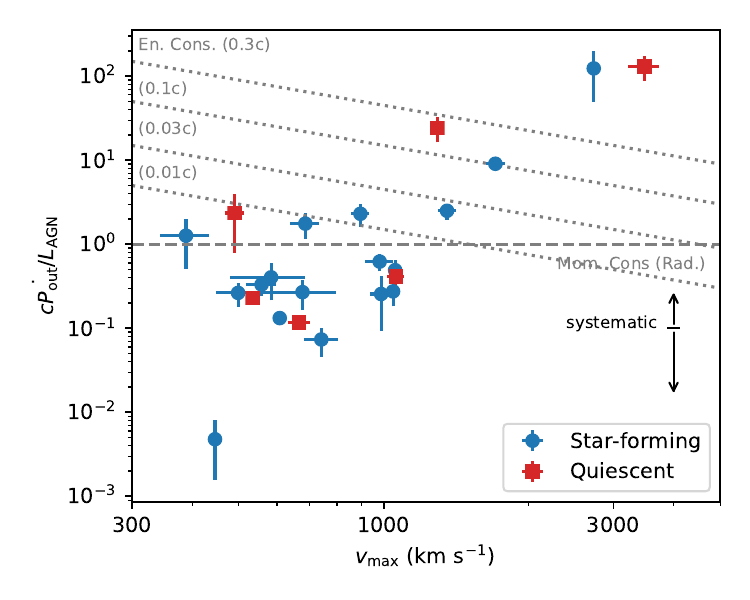}
		\caption{The ratio of the momentum flux carried by the outflows to that produced by the radiation of the AGN versus  $v_\mathrm{max}$. Star-forming galaxies are shown in blue while quiescent galaxies are shown in red. The horizontal dashed line shows the predicted momentum ratio for a momentum-conserving wind driven by radiation pressure on dust, with an IR optical depth of unity. The dotted lines show the predicted momentum ratios for an energy-conserving wind driven by a fast, small scale wind with $v_\mathrm{in} = 0.01c, 0.03c, 0.1c$,~and~$0.3c$.}
		\label{fig:pratio}
\end{figure}

In Figure \ref{fig:pratio} we show the ratio of $\dot{P}_\mathrm{out}$ to $\lagn/c$ versus $v_\mathrm{max}$.
We also show the expected momentum flux ratio for a momentum conserving radiation pressure-driven wind model with $\uptau_\mathrm{IR} = 1$ and for an energy conserving wind model for $v_\mathrm{in} = 0.03 c, 0.1 c$,~and~$0.3 c$.
These velocities are comparable to those observed in the ultrafast outflows in X-rays reported in \citet{fio17}.
The typical observed momentum ratios of the outflows are between 0.1 and 10.
About $40 \%$ of the outflows have momentum ratios above unity, meaning that either an IR optical depth higher than unity is needed for a momentum-conserving wind, or a momentum boost from an energy-conserving wind is required.
Three outflows have momentum ratios within the prediction of an energy-conserving wind with $v_\mathrm{in}$ between $0.01c$ and $0.3c$. 
This shows that for these outflows, they can be driven by small-scale winds similar to typical X-ray ultrafast outflows.
However, the majority of outflows lie below the predicted momentum ratio by energy-conserving winds, such that either these outflows are not strictly energy conserving or some momentum is carried by outflows in other phases, since only ionized outflows are probed here.

Two outflows have momentum ratios higher than the prediction for $v_\mathrm{in} = 0.3c$.
It should be noted that, as $v_\mathrm{max}$ is used to estimate the 3-dimensional velocity  of the outflows from the measured 1-dimensional velocity in general, it may not be strictly accurate for every  outflow depending on the actual outflow geometry in each source.
Moreover, when calculating outflow rates, a single electron density of $n_e = 150 \mathrm{~cm~s}^{-1}$ is assumed for all the outflows in our sample as an average, but the values of $n_e$ in individual sources can vary.
For these two outflows, they possibly have smaller actual velocity than $v_\mathrm{max}$ and/or higher actual electron density than the assumed average, which can lead to overestimation of the momentum ratios and velocities seen in this diagram.

\section{Conclusions}\label{conclusions}

We use data from the complete MOSDEF survey to analyze the rest-frame optical spectra of a sample of 159 X-ray, IR and/or optical AGNs at $1.4 < z < 3.8$.
This dataset triples the previous MOSDEF sample size in \citet{leung17} and is the largest uniform sample of AGNs for outflow studies with coverage of multiple important rest-frame optical emission lines at this redshift to date. 
The AGNs in our sample have bolometric luminosities of $10^{44-47} \mathrm{~erg~s}^{-1}$ and reside in both quiescent galaxies and star-forming galaxies along and across the star-forming main sequence, with stellar masses of $10^{9-11.7} \msun$.
We identify outflows from additional, typically broadened, kinematic components in the \hbeta, [OIII], \halpha ~and [N II] emission line profiles beyond the narrow-line component.
Our main findings are as follows:

\begin{enumerate}
    \item We detect ionized outflows at S/N $>3$ in $17 \%$ of the 159 AGNs. By contrast, using the same analysis procedures, ionized outflows are detected in emission in only $2.5 \%$ out of a mass-matched sample of 1179 inactive MOSDEF galaxies. The $\sim 7$ times higher incidence of outflows in AGNs strongly suggests that these outflows are AGN-driven. (\S\ref{oflw-det}, \S\ref{gal-det})
    
    \item The typical velocity of the outflows is $\approx 940$ \kmps, ranging from $400 - 3500$ \kmps. 22 out of the 31 detected outflows are significantly spatially extended, with sizes extending $0.3-11.0$ kpc, with a median of 4.5 kpc.  These are therefore fast, galaxy-wide outflows. (\S\ref{kin}, \S\ref{ext})
    
    \item In the BPT diagram, the line ratios of the outflowing gas are shifted towards the AGN region compared to the narrow-line region gas. The [N II]/\halpha ~line ratios of the outflowing gas are not correlated with the velocity dispersion, showing no evidence of shock excitation. The outflowing gas is consistent with being photoionized by the AGN. (\S\ref{bpt})
    
    \item The incidence of outflows is largely uniform in stellar mass and SFR of the host galaxy. While the incidence of outflows {\it among all galaxies} increases with stellar mass, the incidence of outflows {\it among detected AGNs} remains constant in stellar mass. The former trend increases at a rate similar to the increasing incidence of AGNs with stellar mass and likely results from the known selection bias of AGNs with stellar mass. The true incidence of outflows is constant with respect to stellar mass. Outflows are detected in both star-forming and quiescent galaxies, reflecting that they exist across different phases of galaxy evolution. (\S\ref{msfr})
    
    \item While the incidence of outflows increases with the [OIII] luminosity of the AGN, it also increases with the S/N in the [OIII] emission line spectrum, as luminosity and S/N are highly correlated. At a given S/N, there is no difference in the distribution of the offset from the mean \loiii ~between AGNs with and without outflows. (\S\ref{loiii})
    
    \item The outflow velocity is correlated with \loiii, following a power law with $\vmax \propto \loiii^{0.27 \pm 0.09}$, consistent with theoretical models of an energy-conserving outflow. No significant correlation is observed between the outflow size and \loiii. As our sample spans $\loiii > 10^{42} \mathrm{~erg~s}^{-1}$, our results support the existence of an upper limit in outflow size of $\sim 5-10$ kpc above this luminosity. The outflow velocity is inversely correlated with the outflow size. This is consistent with an energy-conserving solution of an expanding outflow, while the momentum-conserving solution underestimates the observed velocities. (\S\ref{v-r})
    
    \item The mass outflow rate and kinetic power of the outflow are positively correlated with \lagn. The kinetic coupling efficiency of the outflow ($\dot{E}_\mathrm{out} / \lagn$) is typically $0.01-1 \%$. The AGN are energetically sufficient to drive these outflows, while stellar feedback is insufficient. This suggests that these outflows are very likely AGN driven. There is a weak anti-correlation between mass outflow rate and stellar mass for outflows in star-forming galaxies, while the mass outflow rate in quiescent galaxies deviates upwards of this trend. This suggests that at a given stellar mass, outflows in quiescent galaxies are more powerful than those in star-forming galaxies. (\S\ref{rates-corr}, \S\ref{driver})
    
    \item The mass-loading factor of the outflows is typically between $0.1-1$ and is positively correlated with \lagn ~{\it in star-forming galaxies}. This shows that more luminous AGNs drive more powerful outflows that have higher impact to the host star-forming galaxy, capable of producing mass-loading factors larger than unity at $\lagn \gtrsim 10^{46} \mathrm{~erg~s}^{-1}$. The mass-loading factor {\it in quiescent galaxies} is systematically elevated due to the very low SFR of the host galaxy and can only show whether the outflow may keep the galaxy quenched, but does not indicate the potential role of the outflow in the initial quenching of star formation in the host galaxy. (\S\ref{eta})
    
    \item The momentum flux of the outflows is typically between $0.1-10$ times the momentum flux output from the radiation pressure of the AGN. This ratio is greater than unity for about $40 \%$ of the outflows. For these sources, the outflow can be produced either by a momentum-boost from an energy-conserving wind with an initial velocity of $0.01-0.3 c$ or large optical depths at IR wavelengths due to dust  around the AGN. (\S\ref{mom})
    
\end{enumerate}

Our results provide a census of AGN-driven ionized outflows in the galaxy population at $z \sim 1-3$, but much remains to be understood.
Key observational challenges remain in terms of determining the electron density of the outflows as well as a simultaneous characterization of the multi-phase content of the outflows.
The electron density is crucial in the computation of the outflow rates, but the [S II] emission line ratio  from which it is inferred is typically too weak to be significantly detected in outflows at $z \sim 2$.
The systematic uncertainty in electron density can change the measured outflow rates and mass-loading factors by a factor of a few.
While we find that luminous AGNs are capable of producing ionized outflows with mass-loading factors above unity and can potentially bring substantial impact to the host galaxy, a large scatter exists in this trend and the mass-loading factor is modest in most sources.
Combined with the uncertainty in electron density, it is unclear whether for the majority of the AGNs the ionized outflows alone are sufficient to quench or only regulate the star formation of the host galaxy.
To account for the full impact of the outflows, it is necessary to characterize the multi-phase content of the outflows, including molecular and neutral gas.
Only a few examples of molecular AGN outflows currently exist at $z \sim 2$ \citep[e.g.][]{vay17, bru18, her19}.
Future progress will greatly benefit from high sensitivity observations of the ionized, neutral and molecular content of AGN-driven outflows in representative samples of galaxies at high redshift.
\bigbreak

We thank D. S. Rupke for very useful discussions.
We also thank the anonymous referee for helping to improve the paper.
This work would not have been possible without the 3D-HST collaboration, who provided us with the spectroscopic and photometric catalogs used to select our targets and to derive stellar population parameters.
We acknowledge support from NSF AAG grants AST-1312780, 1312547, 1312764, and 1313171, archival grant AR-13907 provided by NASA through the Space Telescope Science Institute, and grant NNX16AF54G from the NASA ADAP program.
JA acknowledges support from an STFC Ernest Rutherford Fellowship, grant code: ST/P004172/1.
The data presented herein were obtained at the W.M. Keck Observatory, which is operated as a scientific partnership among the California Institute of Technology, the University of California and the National Aeronautics and Space Administration. 
The Observatory was made possible by the generous financial support of the W.M. Keck Foundation. 
The authors wish to recognize and acknowledge the very significant cultural role and reverence that the summit of Mauna Kea has always had within the indigenous Hawaiian community. 
We are most fortunate to have the opportunity to conduct observations from this mountain.

\bibliographystyle{apj}

\begin{thebibliography}{}
\expandafter\ifx\csname natexlab\endcsname\relax\def\natexlab#1{#1}\fi

\bibitem[Aird et al.(2012)]{aird12}
	Aird, J., Coil, A.~L., Moustakas, J., et al.\ 2012, \apj, 746, 90
\bibitem[Aird et al.(2015)]{aird15}
	Aird, J., Coil, A.~L., Georgakakis, A., et al.\ 2015, \mnras, 451, 1892 
\bibitem[Aird et al.(2018)]{aird18} 
	Aird, J., Coil, A.~L., \& Georgakakis, A.\ 2018, \mnras, 474, 1225.
\bibitem[Aird et al.(2019)]{aird19} 
    Aird, J., Coil, A.~L., \& Georgakakis, A.\ 2019, \mnras, 484, 4360
\bibitem[Antonucci(1993)]{ant93}
	Antonucci, R. 1993, \araa, 31, 473
\bibitem[Azadi et al.(2015)]{aza15} 
    Azadi, M., Aird, J., Coil, A.~L., et al.\ 2015, \apj, 806, 187
\bibitem[Azadi et al.(2017)]{aza17} 
	Azadi, M., Coil, A.~L., Aird, J., et al.\ 2017, \apj, 835, 27
\bibitem[Azadi et al.(2018)]{aza18} 
	Azadi, M., Coil, A., Aird, J., et al.\ 2018, ArXiv e-prints , arXiv:1806.08989.
\bibitem[Baldry et al.(2012)]{bal12} 
    Baldry, I.~K., Driver, S.~P., Loveday, J., et al.\ 2012, \mnras, 421, 621.
\bibitem[Baldwin et al.(1981)]{bal81}
	Baldwin, J.~A., Phillips, M.~M., \& Terlevich, R.\ 1981, \pasp, 93, 5
\bibitem[Bae, \& Woo(2016)]{bae16} 
    Bae, H.-J., \& Woo, J.-H.\ 2016, \apj, 828, 97.
\bibitem[Baron, \& Netzer(2019a)]{bar19a} 
    Baron, D., \& Netzer, H.\ 2019, \mnras, 482, 3915
\bibitem[Baron, \& Netzer(2019b)]{bar19b} 
    Baron, D., \& Netzer, H.\ 2019, \mnras, 486, 4290
\bibitem[Behroozi et al.(2013a)]{beh13a} 
    Behroozi, P.~S., Wechsler, R.~H., \& Conroy, C.\ 2013, \apj, 762, L31.
\bibitem[Behroozi et al.(2013b)]{beh13b} 
    Behroozi, P.~S., Wechsler, R.~H., \& Conroy, C.\ 2013, \apj, 770, 57.
\bibitem[Bennert et al.(2002)]{ben02} 
    Bennert, N., Falcke, H., Schulz, H., et al.\ 2002, \apj, 574, L105.
\bibitem[Benson et al.(2003)]{ben03}
	Benson, A. J., Bower, R. G., Frenk, C. S., et al. 2003, \apj, 599, 38
\bibitem[Bouch{\'e} et al.(2010)]{bou10} 
    Bouch{\'e}, N., Dekel, A., Genzel, R., et al.\ 2010, \apj, 718, 1001.
\bibitem[Brusa et al.(2007)]{bru07}
	Brusa, M., Zamorani, G., Comastri, A., et al.\ 2007, \apjs, 172, 353
\bibitem[Brusa (2015)]{bru15}
	Brusa, M., Bongiorno, A., Cresci, G., et al. 2015, \mnras, 446, 2394
\bibitem[Brusa et al.(2018)]{bru18} 
    Brusa, M., Cresci, G., Daddi, E., et al.\ 2018, \aap, 612, A29.
\bibitem[Calzetti et al.(2000)]{cal00}
	Calzetti, D., Armus, L., Bohlin, R.~C., et al.\ 2000, \apj, 533, 682
\bibitem[Cano-D{\'\i}az et al.(2012)]{can12} 
    Cano-D{\'\i}az, M., Maiolino, R., Marconi, A., et al.\ 2012, \aap, 537, L8
\bibitem[Chabrier(2003)]{cha03}
	Chabrier, G.\ 2003, \pasp, 115, 763
\bibitem[Cicone et al.(2014)]{cic14} 
    Cicone, C., Maiolino, R., Sturm, E., et al.\ 2014, \aap, 562, A21.
\bibitem[Coatman et al.(2019)]{coa19} 
    Coatman, L., Hewett, P.~C., Banerji, M., et al.\ 2019, \mnras, 486, 5335
\bibitem[Coil et al.(2015)]{coil15}
	Coil, A.~L., Aird, J., Reddy, N., et al.\ 2015, \apj, 801, 35
\bibitem[Conroy et al.(2009)]{con09}
	Conroy, C., Gunn, J.~E., \& White, M.\ 2009, \apj, 699, 486
\bibitem[Costa et al.(2014)]{cos14} 
    Costa, T., Sijacki, D., \& Haehnelt, M.~G.\ 2014, \mnras, 444, 2355.
\bibitem[Costa et al.(2018)]{cos18} 
    Costa, T., Rosdahl, J., Sijacki, D., et al.\ 2018, \mnras, 473, 4197.
\bibitem[Crenshaw et al.(2003)]{cren03} 
Crenshaw, D.~M., Kraemer, S.~B., \& George, I.~M.\ 2003, Annual Review of Astronomy and Astrophysics, 41, 117.
\bibitem[Croton et al.(2006)]{cro06} 
    Croton, D.~J., Springel, V., White, S.~D.~M., et al.\ 2006, \mnras, 365, 11.
\bibitem[Debuhr et al.(2012)]{deb12}
	Debuhr, J., Quataert, E., \& Ma, C.-P. 2012, \mnras, 420, 2221
\bibitem[Di Matteo et al.(2005)]{dm05}
	Di Matteo, T., Springel, V., \& Hernquist, L.\ 2005, \nat, 433, 604
\bibitem[Donley et al.(2012)]{don12}
	Donley, J.~L., Koekemoer, A.~M., Brusa, M., et al.\ 2012, \apj, 748, 142
\bibitem[Dutton et al.(2010)]{dut10} 
    Dutton, A.~A., van den Bosch, F.~C., \& Dekel, A.\ 2010, \mnras, 405, 1690.
\bibitem[Faucher-Gigu{\`e}re \& Quataert(2012)]{fau12}
	Faucher-Gigu{\`e}re, C.-A., \& Quataert, E.\ 2012, \mnras, 425, 605 
\bibitem[Ferrarese \& Merritt(2000)]{fer00}
	Ferrarese, L., \& Merritt, D. 2000, \apj, 539, L9
\bibitem[Fiore et al.(2017)]{fio17}
    Fiore, F., Feruglio, C., Shankar, F., et al.\ 2017, \aap, 601, A143.
\bibitem[F{\"o}rster Schreiber et al.(2019)]{for18} 
    F{\"o}rster Schreiber, N.~M., {\"U}bler, H., Davies, R.~L., et al.\ 2019, \apj, 875, 21.
\bibitem[Freeman et al.(2019)]{free17} 
    Freeman, W.~R., Siana, B., Kriek, M., et al.\ 2019, \apj, 873, 102.
\bibitem[Gebhardt et al.(2000)]{geb00}
	Gebhardt, K., Bender, R., Bower, G., et al. 2000, \apj, 539, L13
\bibitem[Genzel et al.(2014)]{gen14}
	Genzel, R., F{\"o}rster Schreiber, N.~M., Rosario, D., et al.\ 2014, \apj, 796, 7
\bibitem[Gnedin \& Hollon(2012)]{gne12} 
	Gnedin, N.~Y., \& Hollon, N.\ 2012, \apjs, 202, 13 
\bibitem[Greene et al.(2011)]{gre11} 
    Greene, J.~E., Zakamska, N.~L., Ho, L.~C., et al.\ 2011, \apj, 732, 9
\bibitem[Hainline et al.(2013)]{hain13} 
    Hainline, K.~N., Hickox, R., Greene, J.~E., et al.\ 2013, \apj, 774, 145.
\bibitem[Hainline et al.(2014)]{hain14} 
    Hainline, K.~N., Hickox, R.~C., Greene, J.~E., et al.\ 2014, \apj, 787, 65.
\bibitem[Harrison et al.(2018)]{har18} 
    Harrison, C.~M., Costa, T., Tadhunter, C.~N., et al.\ 2018, Nature Astronomy, 2, 198.
\bibitem[Harrison et al.(2014)]{har14}
	Harrison, C. M., Alexander, D. M., Mullaney, J. R., \& Swinbank, A. M. 2014, \mnras, 441,3306
\bibitem[Harrison et al.(2016)]{har16}
	Harrison, C.~M., Alexander, D.~M., Mullaney, J.~R., et al.\ 2016, \mnras, 456, 1195
\bibitem[Harrison et al.(2012)]{har12}
	Harrison, C.~M., Alexander, D.~M., Swinbank, A.~M., et al.\ 2012, \mnras, 426, 1073 
\bibitem[Heckman \& Best(2014)]{hec14}
	Heckman, T. M., \& Best, P. N. 2014, \araa, 52, 589
\bibitem[Herrera-Camus et al.(2019)]{her19} 
    Herrera-Camus, R., Tacconi, L., Genzel, R., et al.\ 2019, \apj, 871, 37.
\bibitem[Ho et al.(2014)]{ho14} 
Ho, I.-T., Kewley, L.~J., Dopita, M.~A., et al.\ 2014, \mnras, 444, 3894.
\bibitem[Holt et al.(2011)]{hol11} 
    Holt, J., Tadhunter, C.~N., Morganti, R., et al.\ 2011, \mnras, 410, 1527
\bibitem[Hopkins et al.(2006)]{hop06a}
	Hopkins, P.~F., Hernquist, L., Cox, T.~J., et al.\ 2006, \apjs, 163, 1
\bibitem[Hopkins et al.(2008)]{hop08} 
    Hopkins, P.~F., Cox, T.~J., Kere{\v{s}}, D., et al.\ 2008, The Astrophysical Journal Supplement Series, 175, 390.
\bibitem[Hopkins et al.(2012)]{hop12} 
	Hopkins, P.~F., Quataert, E., \& Murray, N.\ 2012, \mnras, 421, 3522 
\bibitem[Husemann et al.(2014)]{huse14} 
    Husemann, B., Jahnke, K., S{\'a}nchez, S.~F., et al.\ 2014, \mnras, 443, 755.
\bibitem[Husemann et al.(2016)]{huse16} 
Husemann, B., Scharw{\"a}chter, J., Bennert, V.~N., et al.\ 2016, \aap, 594, A44.
\bibitem[Kakkad et al.(2018)]{kak18} 
    Kakkad, D., Groves, B., Dopita, M., et al.\ 2018, \aap, 618, A6.
\bibitem[Kang, \& Woo(2018)]{kang18} 
    Kang, D., \& Woo, J.-H.\ 2018, \apj, 864, 124.
\bibitem[Karouzos et al.(2016a)]{kar16a} 
    Karouzos, M., Woo, J.-H., \& Bae, H.-J.\ 2016, \apj, 819, 148
\bibitem[Karouzos et al.(2016b)]{kar16b} 
    Karouzos, M., Woo, J.-H., \& Bae, H.-J.\ 2016, \apj, 833, 171
\bibitem[Kauffmann et al.(2003)]{kau03}
	Kauffmann, G., Heckman, T.~M., Tremonti, C., et al.\ 2003, \mnras, 346, 1055
\bibitem[Kauffmann \& Heckman(2009)]{kau09} 
	Kauffmann, G., \& Heckman, T.~M.\ 2009, \mnras, 397, 135
\bibitem[Kaviraj et al.(2017)]{kav17} 
Kaviraj, S., Laigle, C., Kimm, T., et al.\ 2017, \mnras, 467, 4739.
\bibitem[Kelly(2007)]{kel07} 
    Kelly, B.~C.\ 2007, \apj, 665, 1489.
\bibitem[Kennicutt(1998)]{ken98} 
	Kennicutt, R.~C., Jr.\ 1998, \araa, 36, 189 
\bibitem[Kewley et al.(2013)]{kew13} 
	Kewley, L.~J., Dopita, M.~A., Leitherer, C., et al.\ 2013a, \apj, 774, 100 
\bibitem[King et al.(2011)]{kin11} 
    King, A.~R., Zubovas, K., \& Power, C.\ 2011, \mnras, 415, L6.
\bibitem[Kriek et al.(2009)]{kri09}
	Kriek, M., van Dokkum, P.~G., Labb{\'e}, I., et al.\ 2009, \apj, 700, 221 
\bibitem[Kriek et al.(2015)]{kri15}
	Kriek, M., Shapley, A.~E., Reddy, N.~A., et al.\ 2015, \apjs, 218, 15
\bibitem[Laird et al.(2009)]{lai09}
	Laird, E.~S., Nandra, K., Georgakakis, A., et al.\ 2009, \apjs, 180, 102
\bibitem[Leung et al.(2017)]{leung17} 
	Leung, G.~C.~K., Coil, A.~L., Azadi, M., et al.\ 2017, \apj, 849, 48.
\bibitem[Liu et al.(2013a)]{liu13a}
	Liu, G., Zakamska, N. L., Greene, J. E., et al. 2013, \mnras, 430, 2327
\bibitem[Liu et al.(2013b)]{liu13b}
	Liu, G., Zakamska, N.~L., Greene, J.~E., Nesvadba, N.~P.~H., \& Liu, X.\ 2013, \mnras, 436, 2576 
\bibitem[Luo et al.(2010)]{luo10}
	Luo, B., Brandt, W.~N., Xue, Y.~Q., et al.\ 2010, \apjs, 187, 560
\bibitem[Madau et al.(1996)]{mad96} 
    Madau, P., Ferguson, H.~C., Dickinson, M.~E., et al.\ 1996, \mnras, 283, 1388.
\bibitem[Madau, \& Dickinson(2014)]{mad14} 
    Madau, P., \& Dickinson, M.\ 2014, Annual Review of Astronomy and Astrophysics, 52, 415.
\bibitem[Magorrian et al.(1998)]{mag98}
	Magorrian, J., Tremaine, S., Richstone, D., et al.\ 1998, \aj, 115, 2285
\bibitem[Markwardt(2009)]{mar09} 
	Markwardt, C.~B.\ 2009, Astronomical Data Analysis Software and Systems XVIII, 411, 251
\bibitem[Mel{\'e}ndez et al.(2014)]{mel14} 
	Mel{\'e}ndez, M., Heckman, T.~M., Mart{\'{\i}}nez-Paredes, M., Kraemer, S.~B., \& Mendoza, C.\ 2014, \mnras, 443, 1358 
\bibitem[McElroy et al.(2015)]{mce15} McElroy, R., Croom, S.~M., Pracy, M., et al.\ 2015, \mnras, 446, 2186.
\bibitem[McLean et al.(2010)]{mcl10}
	McLean, I.~S., Steidel, C.~C., Epps, H., et al.\ 2010, \procspie, 7735, 77351E-77351E-12
\bibitem[McLean et al.(2012)]{mcl12}
	McLean, I.~S., Steidel, C.~C., Epps, H.~W., et al.\ 2012, \procspie, 8446, 84460J
\bibitem[Mingozzi et al.(2019)]{min19} 
    Mingozzi, M., Cresci, G., Venturi, G., et al.\ 2019, \aap, 622, A146
\bibitem[Mullaney et al.(2013)]{mul13}
	Mullaney, J. R., Alexander, D. M., Fine, S., et al. 2013, \mnras, 433, 622
\bibitem[Nandra et al.(2015)]{nan15}
	Nandra, K., Laird, E.~S., Aird, J.~A., et al.\ 2015, \apjs, 220, 10
\bibitem[Nesvadba et al.(2008)]{nes08} 
	Nesvadba, N.~P.~H., Lehnert, M.~D., De Breuck, C., Gilbert, A.~M., \& van Breugel, W.\ 2008, \aap, 491, 407 
\bibitem[Nesvadba et al.(2017)]{nes17a} 
	Nesvadba, N.~P.~H., De Breuck, C., Lehnert, M.~D., Best, P.~N., \& Collet, C.\ 2017, \aap, 599, A123 
\bibitem[Netzer(2015)]{net15}
	Netzer, H. 2015, \araa, 53, 365
\bibitem[Osterbrock \& Ferland(2006)]{ost06}
	Osterbrock, D.~E., \& Ferland, G.~J.\ 2006, Astrophysics of gaseous nebulae and active galactic nuclei, 2nd.~ed.~by D.E.~Osterbrock and G.J.~Ferland.~Sausalito, CA: University Science Books
\bibitem[Perna et al.(2015)]{per15}
	Perna, M., Brusa, M., Cresci, G., et al.\ 2015, \aap, 574, A82
\bibitem[Perna et al.(2017a)]{per17a} 
    Perna, M., Lanzuisi, G., Brusa, M., et al.\ 2017, \aap, 603, A99
\bibitem[Perna et al.(2017b)]{per17b} 
    Perna, M., Lanzuisi, G., Brusa, M., et al.\ 2017, \aap, 606, A96.
\bibitem[Pillepich et al.(2018)]{phi18} 
Pillepich, A., Springel, V., Nelson, D., et al.\ 2018, \mnras, 473, 4077.
\bibitem[Polletta et al.(2007)]{pol07} 
	Polletta, M., Tajer, M., Maraschi, L., et al.\ 2007, \apj, 663, 81.
\bibitem[Reddy et al.(2015)]{red15} 
	Reddy, N.~A., Kriek, M., Shapley, A.~E., et al.\ 2015, \apj, 806, 259 
\bibitem[Rose et al.(2018)]{ros18} 
    Rose, M., Tadhunter, C., Ramos Almeida, C., et al.\ 2018, \mnras, 474, 128
\bibitem[Rupke et al.(2005)]{rup05}
	Rupke, D.~S., Veilleux, S., \& Sanders, D.~B.\ 2005, \apjs, 160, 115
\bibitem[Rupke, \& Veilleux(2013)]{rup13} 
    Rupke, D.~S.~N., \& Veilleux, S.\ 2013, \apj, 768, 75.

\bibitem[Rupke et al.(2017)]{rup17} 
    Rupke, D.~S.~N., G{\"u}ltekin, K., \& Veilleux, S.\ 2017, \apj, 850, 40.
\bibitem[Schaye et al.(2015)]{sch15} 
    Schaye, J., Crain, R.~A., Bower, R.~G., et al.\ 2015, \mnras, 446, 521.
\bibitem[Schmitt et al.(2003)]{sch03} 
    Schmitt, H.~R., Donley, J.~L., Antonucci, R.~R.~J., et al.\ 2003, \apj, 597, 768.
\bibitem[Sharples et al.(2013)]{sha13} 
    Sharples, R., Bender, R., Agudo Berbel, A., et al.\ 2013, The Messenger, 151, 21.
\bibitem[Shivaei et al.(2015)]{shiv15}
	Shivaei, I., Reddy, N.~A., Shapley, A.~E., et al.\ 2015, \apj, 815, 98 
\bibitem[Shivaei et al.(2018)]{shiv18} 
    Shivaei, I., Reddy, N.~A., Siana, B., et al.\ 2018, \apj, 855, 42.
\bibitem[Silva et al.(2004)]{sil04} 
	Silva, L., Maiolino, R., \& Granato, G.~L.\ 2004, \mnras, 355, 973.
\bibitem[Skelton et al.(2014)]{skel14}
	Skelton, R.~E., Whitaker, K.~E., Momcheva, I.~G., et al.\ 2014, \apjs, 214, 24
\bibitem[Steidel et al.(2010)]{stei10} 
	Steidel, C.~C., Erb, D.~K., Shapley, A.~E., et al.\ 2010, \apj, 717, 289 
\bibitem[Sun et al.(2017)]{sun17} 
    Sun, A.-L., Greene, J.~E., \& Zakamska, N.~L.\ 2017, \apj, 835, 222.
\bibitem[Thompson et al.(2015)]{thom15}
	Thompson, T.~A., Fabian, A.~C., Quataert, E., \& Murray, N.\ 2015, \mnras, 449, 147 
\bibitem[Vayner et al.(2017)]{vay17} 
    Vayner, A., Wright, S.~A., Murray, N., et al.\ 2017, \apj, 851, 126.
\bibitem[Veilleux \& Osterbrock(1987)]{vei87}
	Veilleux, S., \& Osterbrock, D.~E.\ 1987, \apjs, 63, 295
\bibitem[Veilleux et al.(2013)]{vei13} 
    Veilleux, S., Mel{\'e}ndez, M., Sturm, E., et al.\ 2013, \apj, 776, 27.
\bibitem[Woo et al.(2016)]{woo16}
	Woo, J.-H., Bae, H.-J., Son, D., \& Karouzos, M.\ 2016, \apj, 817, 108 
\bibitem[Yang et al.(2018)]{yang18} 
    Yang, G., Brandt, W.~N., Vito, F., et al.\ 2018, \mnras, 475, 1887
\bibitem[Zakamska et al.(2006)]{zak06} 
    Zakamska, N.~L., Strauss, M.~A., Krolik, J.~H., et al.\ 2006, \aj, 132, 1496
\bibitem[Zakamska \& Greene(2014)]{zak14}
	Zakamska, N.~L., \& Greene, J.~E.\ 2014, \mnras, 442, 784

\end{thebibliography}

\appendix

\section{Emission line spectra and tabulated data}

In this Appendix, we we provide the emission line spectra of the AGN with detected outflows and tabulated data of the AGN, host galaxy and outflow properties in our sample.

\capstartfalse
\begin{deluxetable}{lccccr@{}lr@{}lr@{}lr@{}l}[!h]
\centering
\tablecaption{AGN, host galaxy and outflow properties}
\tablehead{
\colhead{ID\tablenotemark{a}} & 
\colhead{$z$} &
\colhead{$\log(\loiii)$}&
\colhead{$M_*$} &
\colhead{SFR}&
\multicolumn{2}{c}{\vmax}&
\multicolumn{2}{c}{$r_{10,\mathrm{[OIII]}}$}&
\multicolumn{2}{c}{$\dot{M}$}&
\multicolumn{2}{c}{$\dot{E}$}
\\
\colhead{}&
\colhead{}&
\colhead{($\mathrm{erg~s}^{-1}$)}&
\colhead{($\msun$)}&
\colhead{($\msun~\mathrm{yr}^{-1}$)}&
\multicolumn{2}{c}{(\kmps)}&
\multicolumn{2}{c}{(kpc)}&
\multicolumn{2}{c}{($\msun~\mathrm{yr}^{-1}$)}&
\multicolumn{2}{c}{($\mathrm{erg~s}^{-1}$)}
}
\startdata
1606	&	2.475	&	43.04	&	10.77	&	105	& 	-2605	$\pm$ &	194	& 	1.66	$\pm$ &	1.33	& 	7.1	$\pm$ &	54.7	 & $(	1.5	\pm$ & $	11.7	) \times 10 ^{	43	}$ \\
5095	&	2.295	&	42.62	&	10.12	&	355	& 	-1706	$\pm$ &	78	& 	2	$\pm$ &	0.88	& 	60.9	$\pm$ &	7.6	 & $(	5.6	\pm$ & $	1.0	) \times 10 ^{	43	}$ \\
5130	&	2.377	&	42.88	&	10.96	&	7	& 	-534	$\pm$ &	11	& 	2.02	$\pm$ &	1.74	& 	8.8	$\pm$ &	0.6	 & $(	7.9	\pm$ & $	0.7	) \times 10 ^{	41	}$ \\
5224	&	2.151	&	42.65	&	10.81	&	$<1$	& 	-490	$\pm$ &	23	& 	0.28	$\pm$ &	0.17	& 	62.9	$\pm$ &	41.8	 & $(	4.8	\pm$ & $	3.2	) \times 10 ^{	42	}$ \\
6743	&	2.487	&	43.79	&	9.66	&	155	& 	-2731	$\pm$ &	46	& 	0.06	$\pm$ &	0.03	& 	8050.4	$\pm$ &	4839.2	 & $(	1.9	\pm$ & $	1.1	) \times 10 ^{	46	}$ \\
8388	&	2.198	&	42.74	&	11.03	&	$<1$	& 	-1513	$\pm$ &	161	& 	0.05	$\pm$ &	0.34	& 	55.8	$\pm$ &	60.1	 & $(	4.0	\pm$ & $	4.5	) \times 10 ^{	43	}$ \\
10769-1	&	2.103	&	42.58	&	11.18	&	219	& 	-742	$\pm$ &	60	& 	8.83	$\pm$ &	0.89	& 	1.2	$\pm$ &	0.4	 & $(	2.1	\pm$ & $	0.9	) \times 10 ^{	41	}$ \\
10769-2	&	2.103	&	42.58	&	11.18	&	219	& 	-446	$\pm$ &	11	& 	11	$\pm$ &	0.3	& 	0.1	$\pm$ &	0.1	 & $(	8.1	\pm$ & $	5.5	) \times 10 ^{	39	}$ \\
11487	&	3.408	&	43.62	&	10.36	&	6	& 	-1807	$\pm$ &	141	& 	0.09	$\pm$ &	0.1	& 	1973.1	$\pm$ &	2423.2	 & $(	2.0	\pm$ & $	2.5	) \times 10 ^{	45	}$ \\
12302	&	2.276	&	42.75	&	10.10	&	170	& 	-981	$\pm$ &	66	& 	0.99	$\pm$ &	0.37	& 	11.0	$\pm$ &	2.5	 & $(	3.3	\pm$ & $	1.0	) \times 10 ^{	42	}$ \\
12615	&	3.179	&	43.53	&	10.86	&	51	& 	1056	$\pm$ &	35	& 	4.5	$\pm$ &	0.37	& 	26.9	$\pm$ &	9.0	 & $(	9.5	\pm$ & $	3.3	) \times 10 ^{	42	}$ \\
13429	&	3.475	&	43.36	&	10.80	&	87	& 	-816	$\pm$ &	135	& 	0.97	$\pm$ &	2.58	& 	78.5	$\pm$ &	212.6	 & $(	1.6	\pm$ & $	4.5	) \times 10 ^{	43	}$ \\
14596	&	2.447	&	42.84	&	11.63	&	$<1$	& 	-3481	$\pm$ &	257	& 	1.34	$\pm$ &	0.36	& 	768.8	$\pm$ &	251.0	 & $(	2.9	\pm$ & $	1.1	) \times 10 ^{	45	}$ \\
15359	&	1.594	&	42.92	&	10.59	&	2	& 	-667	$\pm$ &	35	& 	4.58	$\pm$ &	0.47	& 	4.4	$\pm$ &	0.7	 & $(	6.2	\pm$ & $	1.4	) \times 10 ^{	41	}$ \\
17664	&	2.187	&	42.65	&	10.77	&	26	& 	-678	$\pm$ &	117	& 	5.57	$\pm$ &	1.67	& 	5.4	$\pm$ &	1.9	 & $(	7.9	\pm$ & $	4.7	) \times 10 ^{	41	}$ \\
17754	&	2.297	&	42.30	&	11.24	&	35	& 	-498	$\pm$ &	49	& 	2.63	$\pm$ &	4.71	& 	3.2	$\pm$ &	1.0	 & $(	2.5	\pm$ & $	1.0	) \times 10 ^{	41	}$ \\
19082	&	2.487	&	42.27	&	11.20	&	158	& 	-583	$\pm$ &	103	& 	9.21	$\pm$ &	4.14	& 	3.6	$\pm$ &	1.6	 & $(	3.8	\pm$ & $	2.5	) \times 10 ^{	41	}$ \\
21290	&	2.215	&	43.36	&	9.45	&	6	& 	-1046	$\pm$ &	36	& 	3.02	$\pm$ &	0.91	& 	18.4	$\pm$ &	6.2	 & $(	6.4	\pm$ & $	2.2	) \times 10 ^{	42	}$ \\
21492	&	2.472	&	43.53	&	11.20	&	1	& 	-1292	$\pm$ &	45	& 	0.31	$\pm$ &	0.1	& 	1805.5	$\pm$ &	593.1	 & $(	9.5	\pm$ & $	3.3	) \times 10 ^{	44	}$ \\
22995	&	2.468	&	43.01	&	11.11	&	21	& 	-988	$\pm$ &	49	& 	2.45	$\pm$ &	1.06	& 	7.9	$\pm$ &	5.0	 & $(	2.4	\pm$ & $	1.6	) \times 10 ^{	42	}$ \\
24192	&	2.243	&	42.92	&	9.63	&	63	& 	-1353	$\pm$ &	59	& 	2.22	$\pm$ &	1.16	& 	44.4	$\pm$ &	9.6	 & $(	2.6	\pm$ & $	0.6	) \times 10 ^{	43	}$ \\
26009	&	3.434	&	43.86	&	10.13	&	78	& 	-894	$\pm$ &	38	& 	1.35	$\pm$ &	0.35	& 	361.3	$\pm$ &	106.0	 & $(	9.1	\pm$ & $	2.9	) \times 10 ^{	43	}$ \\
26304	&	1.632	&	42.30	&	10.72	&	23	& 	-557	$\pm$ &	41	& 	0.53	$\pm$ &	0.47	& 	3.7	$\pm$ &	1.0	 & $(	3.6	\pm$ & $	1.2	) \times 10 ^{	41	}$ \\
30014	&	2.291	&	41.55	&	10.36	&	25	& 	-388	$\pm$ &	45	& 	1.89	$\pm$ &	1.32	& 	3.4	$\pm$ &	2.0	 & $(	1.6	\pm$ & $	1.1	) \times 10 ^{	41	}$ \\
30274	&	2.225	&	43.19	&	10.47	&	355	& 	-1432	$\pm$ &	68	& 	1.77	$\pm$ &	3.3	& 	81.8	$\pm$ &	152.8	 & $(	5.3	\pm$ & $	9.9	) \times 10 ^{	43	}$ \\
32856	&	2.270	&	43.15	&	10.76	&	3	& 	1060	$\pm$ &	43	& 	2.83	$\pm$ &	0.8	& 	16.2	$\pm$ &	5.1	 & $(	5.7	\pm$ & $	1.9	) \times 10 ^{	42	}$ \\
33691	&	2.243	&	42.35	&	10.32	&	8	& 	-687	$\pm$ &	47	& 	4.38	$\pm$ &	2.3	& 	15.5	$\pm$ &	5.2	 & $(	2.3	\pm$ & $	0.9	) \times 10 ^{	42	}$ \\
38195	&	1.487	&	42.72	&	9.64	&	129	& 	-608	$\pm$ &	9	& 	8.67	$\pm$ &	0.44	& 	3.4	$\pm$ &	0.2	 & $(	4.0	\pm$ & $	0.3	) \times 10 ^{	41	}$ 
\enddata
\tablenotetext{a}{3D-HST v4 catalog ID.}
\end{deluxetable}

\begin{figure*}[!htbp]
	\centering
		\figurenum{A1}
		\includegraphics[width=0.95\textwidth]{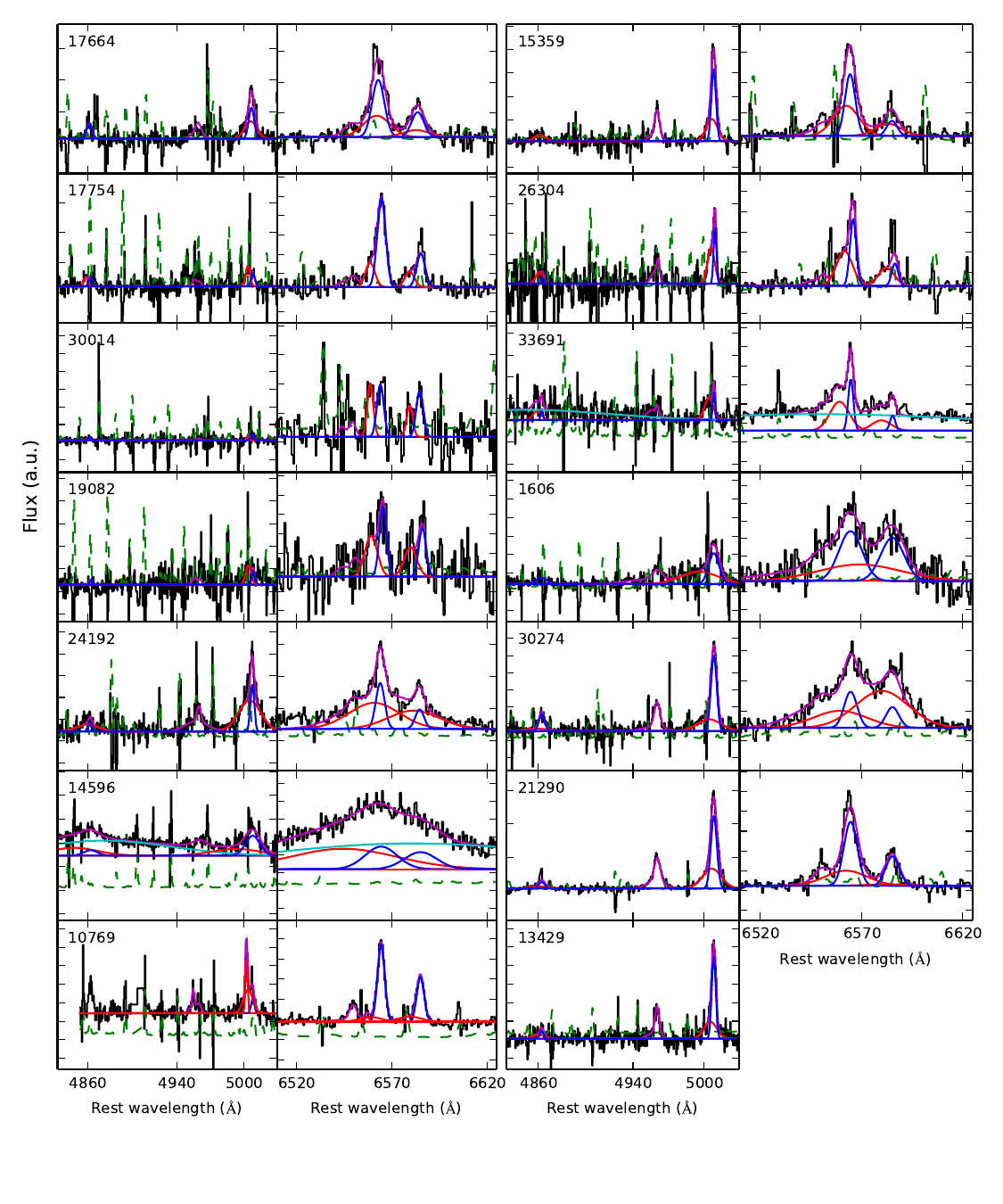}
		\caption{Emission line spectra of the AGN with outflows. For each source, the left panel shows the \hbeta ~and [OIII] emission lines while the right panel shows the [N II] and \halpha ~emission lines. The black line shows the observed spectrum and the green dotted line shows the error spectrum. The magenta line shows the best-fit model, while the blue and red lines shows the best-fit narrow-line and outflow components, respectively. Note that the wavelength ranges of the [OIII] and \halpha ~panels are slightly different, so that the widths of the emission lines in the two panels may appear different.}
\end{figure*}

\begin{figure*}[!htbp]
	\centering
		\figurenum{A1}
		\includegraphics[width=0.95\textwidth]{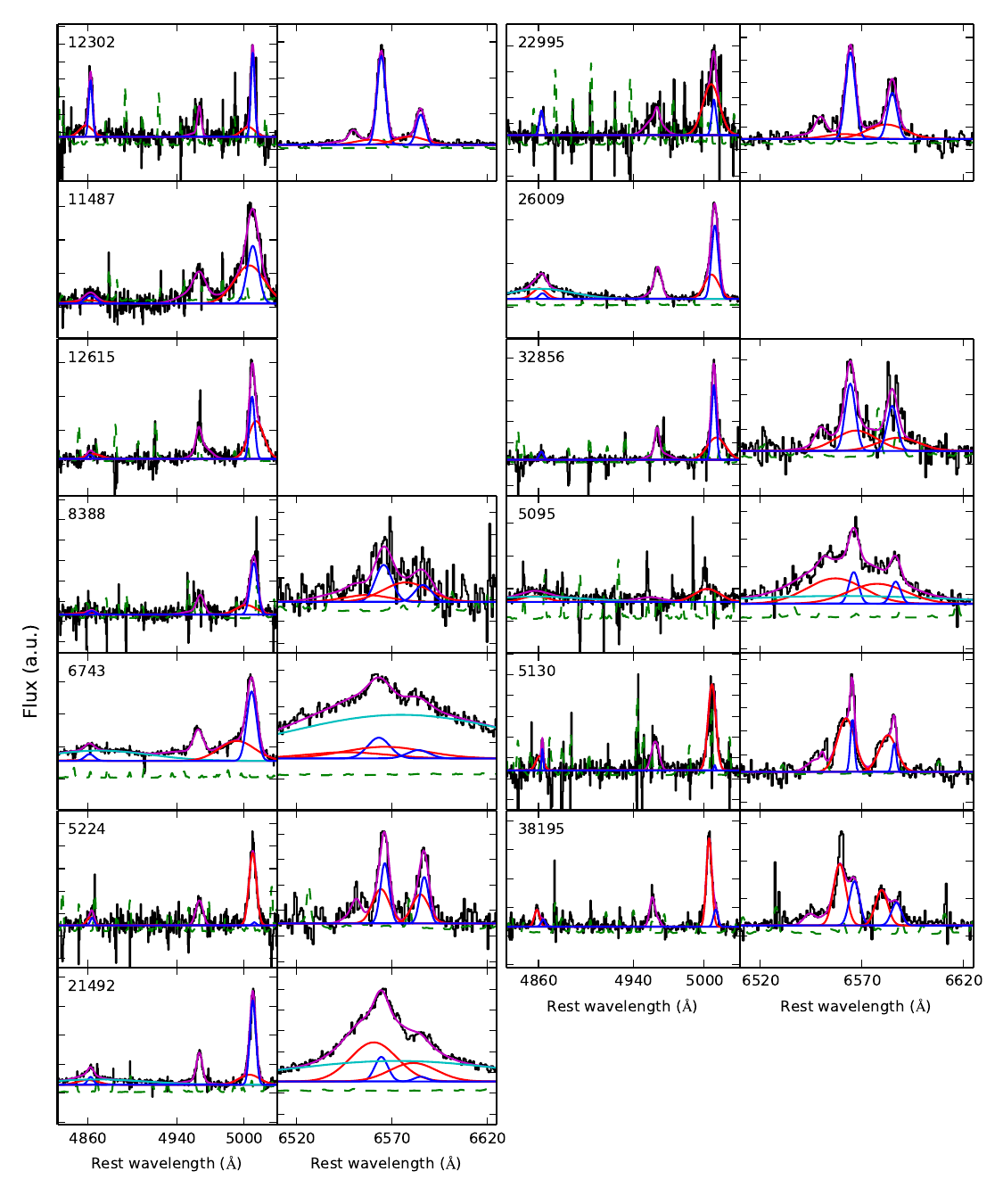}
		\caption{continued}
\end{figure*}

\end{document}